\documentclass[aps,twocolumn,floatfix,altaffilletter,superscriptaddress,preprintnumbers,
               tightenlines,showpacs,showkeys,notitlepage,nofootinbib,
               longbibliography]{revtex4-2}

\usepackage{amsmath,amssymb} % Maths
\usepackage{graphicx} % Figures
\usepackage[export]{adjustbox} % Vertical alignment of figures
\usepackage{url} % URL formatting
\usepackage[dvipsnames]{xcolor} % Colors
\usepackage{siunitx} % Proper unit formating
\usepackage{orcidlink} % for ORCIDs behin author names
\usepackage{tabularx} % Tables
\usepackage{titlesec}
\usepackage{microtype}
\usepackage{dcolumn}% Align table columns on decimal point
\usepackage{bm,bbm}% bold math

\allowdisplaybreaks

\makeatletter
\def\thesection{\arabic{section}}%
\def\p@section{}%
\def\thesubsection{\thesection.\arabic{subsection}}%
\def\p@subsection{}%
\def\thesubsubsection{\thesubsection.\arabic{subsubsection}}%
\def\p@subsubsection{}%
\def\appendix{%
    \par
    \setcounter{section}\z@
    \setcounter{subsection}\z@
    \setcounter{subsubsection}\z@
    \def\thesubsection{\thesection.\arabic{subsection}}%
    \def\thesubsubsection{\thesubsection.\arabic{subsubsection}}%
    \def\p@subsection{}%
    \def\p@subsubsection{}%
    \@addtoreset{equation}{section}%
    \def\theequation@prefix{\thesection}%
    \addtocontents{toc}{\protect\appendix}%
    \@ifstar{%
    \def\thesection{\unskip}%
    \def\theequation@prefix{A.}%
    }{%
    \def\thesection{\Alph{section}}%
    }%
}%
\makeatother

\titleformat*{\section}{\centering\bfseries\MakeUppercase}
\titlelabel{\thetitle\quad}
\titleformat*{\paragraph}{\bfseries}
\titlespacing*{\paragraph}{0pt}{3.25ex plus 1ex minus .2ex}{1em}
\titlelabel{\thetitle\quad}  % remove dot behind section number

\usepackage{hyperref} % References
\usepackage[capitalize]{cleveref} % Nice referencing
\newcommand\myshade{80} % Reference coloring
\colorlet{mylinkcolor}{ForestGreen}
\colorlet{mycitecolor}{Red}
\colorlet{myurlcolor}{violet}
\hypersetup{
  linkcolor  = mylinkcolor!\myshade!black,
  citecolor  = mycitecolor!\myshade!black,
  urlcolor   = myurlcolor!\myshade!black,
  colorlinks = true
}

\renewcommand{\vec}[1]{{\mathbf{#1}}}
\newcommand{\bra}[1]{\ensuremath{\langle #1 |}}   % Bra vector
\newcommand{\ket}[1]{\ensuremath{| #1 \rangle}}   % Ket vector
\newcommand{\amp}[3]{\ensuremath{\left\langle #1 \left| #2%
                     \right| #3 \right\rangle}}  % QM amplitude
\newcommand{\sprod}[2]{\ensuremath{\left\langle #1 |%
                     #2 \right\rangle}}  % QM scalar product
\newcommand{\tr}{\text{tr}}

\renewcommand{\Re}{\operatorname{Re}}

\begin{document}

\title{Entanglement in Collective Neutrino Oscillations on a Quantum Computer}

\newcommand{\utoronto}{Department of Physics, University of Toronto, Toronto, ON M5S 1A7, Canada}
\newcommand{\cern}{CERN, 1211 Geneva 23, Switzerland}
\newcommand{\jgu}{PRISMA Cluster of Excellence \& Mainz Institute for
                  Theoretical Physics, \\
                  Johannes Gutenberg University, 55099 Mainz, Germany}
\newcommand{\tifr}{Tata Institute of Fundamental Research, Homi Bhabha Road, Colaba, Mumbai 400005, India}

\author{Katarina Bleau \orcidlink{0000-0003-0600-4996}\,}
\email{kbleau@uni-mainz.de}
\affiliation{\jgu}

\author{Nikolina Ilic \orcidlink{0000-0003-0105-7634}\,}
\email{nikolina.ilic@cern.ch}
\affiliation{\utoronto}

\author{Joachim Kopp \orcidlink{0000-0003-0600-4996}\,}
\email{jkopp@cern.ch}
\affiliation{\jgu}

\author{Ushak Rahaman \orcidlink{0000-0003-0419-2970}\,}
\email{ushak.rahaman@cern.ch}
\affiliation{\utoronto}
\affiliation{\tifr}

\author{Xin Yue Yu \orcidlink{0000-0002-6976-2703}\,}
\email{xyz.yu@mail.utoronto.ca}
\affiliation{\utoronto}

\date{\today}
\preprint{MITP-26-059} 

%=============================================================================

\begin{abstract}
\noindent
Simulations on a quantum computer have been propounded as a next-generation tool for studying neutrino flavour oscillations in dense environments like supernovae or neutron star mergers. However, quantum advantage hinges on the presence of quantum entanglement in the neutrino ensemble, and it has been argued that such entanglement becomes negligible in the limit of large neutrino number density. Here, we address this question by explicitly simulating the multiparticle dynamics of entangled neutrino ensembles with a varying number of particles, $N$, on both classical and quantum hardware, and comparing our results to those obtained with conventional ``mean-field'' methods based on Boltzmann-type kinetic equations. For simple toy systems, we demonstrate both analytically and numerically that, indeed, the predictions of full multiparticle simulations agree with those of mean-field methods in the large-$N$ limit. Along the way, we develop novel algorithms for the efficient simulation of entangled neutrino gases on both classical and quantum hardware. We then discuss physical situations in which an entangled neutrino system \emph{cannot} be described through mean-field dynamics alone and quantum entanglement therefore becomes phenomenologically relevant. We comment on the conditions under which these scenarios may be realized in supernova cores or neutron star mergers. 
\end{abstract}

%=============================================================================
\maketitle
%=============================================================================

%=============================================================================
\section{Introduction}
\label{sec:intro}
%=============================================================================

Neutrinos are an ideal laboratory for studying the intricacies of quantum mechanics, given that their extremely weak interactions allow them to maintain quantum properties like superpositions and entanglement for extended periods of time. An interesting question in this context is under which circumstances the flavours of different neutrinos can become entangled, and whether this entanglement can have observable consequences. In fact, it has been argued that ultradense environments like supernova cores, neutron star mergers, or the early Universe may provide conditions where such entanglement occurs \cite{%
  Bell:2003mg,      % Entanglement-induced speedup of collisional flavour equilibration
  Cervia:2019res,   % Entanglement growth and departures from mean-field behaviour
  Rrapaj:2019pxz,   % Exact solutions reveal differences between multiparticle and mean-field
  Roggero:2021asb,  % bipolar osc. from many-body correlations; tensor-networks
  Patwardhan:2021rej, % enhanced neutrino entanglement near spectral splits
  Xiong:2021evk,    % deviations from MFT due to many-body decoherence
  Roggero:2022hpy,  % entanglement in fast flavour conversion; scaling with N
  Martin:2023ljq}.  % many-body dynamics in N=16 system, deviations from MFT

In these environments, neutrino densities are so high that neutrino--neutrino coherent forward scattering \cite{Mikheyev:1985zog, Wolfenstein:1977ue}, wherein the neutrino flavour may change while momenta remain unaltered, becomes non-negligible \cite{Pantaleone:1992eq, Pantaleone:1992xh, Qian:1994wh, Samuel:1995ri, Pastor:2002we, Balantekin:2004ug}. This leads to a plethora of novel, ``collective'', effects \cite{Duan:2010bg, Mirizzi:2015eza, Sen:2024fxa, Johns:2025mlm}, an understanding of which is vital in the context of supernova neutrino observations whose goal is to infer the conditions in the supernova core (before oscillations) from the neutrino spectra, ``light'' curves (luminosity as a function of time), and the flavour composition observed upon reaching Earth \cite{Qian:1995ua, Qian:1993dg}. In addition, the neutrino flavour distribution plays a pivotal role in the supernova explosion mechanism itself and in the associated nucleosynthesis processes \cite{Pruet:2005qd, Arcones:2006uq, Fuller:1995ih}. To this day, however, our understanding of how neutrinos change flavour on their way out of a supernova is fragmented. The problem is usually formulated using Boltzmann-type quantum kinetic equations (``mean-field approach'') \cite{Sigl:1993ctk}, but solving these equations for systems that go beyond simple toy models has proven to be a formidable challenge. To describe conditions in a real supernova, one is facing a multiscale problem in six phase space dimensions, and with a background of ordinary matter that is itself highly asymmetric and turbulent. Consequently, a complete understanding of supernova neutrino flavour evolution remains elusive.

Quantum entanglement, which is not captured by quantum kinetic equations, would add a further layer of complications, and would pose a formidable computational challenge, given that the size of the relevant Hilbert space scales exponentially with the number $N$ of simulated neutrinos. Collectively oscillating neutrino systems with entanglement may therefore call for simulations on a \emph{quantum computer}, which would reduce the computational complexity to polynomial in $N$. Pioneering work on this topic has been carried out in refs.~\cite{
    Hall:2021rbv,       % Roggero et al., collective nu oscillations on a quantum computer
    Yeter-Aydeniz:2021olz, % collective nu osc on a quantum computer
    Jha:2021itm,        % Hyderabad group
    Amitrano:2022yyn,   % Roggero et al., Trapped-ion quantum simulation
    Balantekin:2023qvm, % Balantekin et al., QIS / quantum simulation in neutrino physics
    Siwach:2023wzy,     % Balantekin et al., hybrid quantum-classical algorithm
    Turro:2024shh,      % qubits and qutrits
    Chernyshev:2024pqy, % Savage et al., quantum magic in the neutrino sector
    Singh:2024vpu,      % IIT Mohali group, simulations on NMR quantum processor
    Spagnoli:2025etu,   % Roggero et al., 3-flavors, qutrits
    Tripathi:2025cok,   % Mumbai group
    Bleau:2026iiq,      % us :-)
    Heimsoth:2026ozm}.  % Balantekin et al.: hybrid quantum/classical w/ qutrits
However, while there is a clear consensus that entanglement is phenomenologically important in the few-particle toy models that can be studied on current quantum hardware, there are arguments suggesting that it becomes phenomenologically irrelevant at the very high neutrino densities, $n_\nu \sim \SI{e30}{cm^{-3}}$, realized in supernovae and similar environments \cite{Friedland:2003dv, Friedland:2003eh, Shalgar:2023ooi, Johns:2023ewj}. For some toy systems, this can be shown explicitly (see e.g.\ Refs.~\cite{Friedland:2003dv, Friedland:2003eh} and \cref{sec:qke-emergence} below). Moreover, in a three-dimensional system like a supernova or neutron star merger, with neutrinos streaming in all directions, the probability that any pair of neutrinos gets more than one chance to interact is negligible. This ``once-in-a-lifetime encounter'' argument \cite{Kost:2024esc, Kost:2025vyt} leads to the conjecture that entanglement cannot build up efficiently in realistic environments. Both arguments would suggest that quantum kinetic equations provide a sufficiently accurate description in practice. Overall, the quantum kinetic (Boltzmann) approach to collective neutrino oscillations has not been proven wrong, but possible shortcomings due to its inability to capture entanglement have not been ruled out either.

The goals of the present work are twofold: first, we aim to study explicitly the emergence of Boltzmann-like ``mean-field'' behaviour in fully entangled systems of self-interacting neutrinos in the large-$N$ limit. We will do so using analytical calculations, multiparticle simulations on classical hardware, and simulations on superconducting quantum processors. Second, we will introduce neutrino systems in which the preceding arguments do not hold and quantum entanglement among neutrinos becomes phenomenologically relevant. We will discuss under which conditions these systems may be realized in nature.

The paper is structured as follows: in \cref{sec:qke-vs-multiparticle} we first review the quantum kinetic approach to collective neutrino oscillations (\cref{sec:qke-formalism}). For specific systems consisting of just two neutrino modes with similar numbers of neutrinos in each (\cref{sec:bipolar}), we then argue analytically that quantum kinetic behaviour emerges from a full multiparticle treatment including entanglement in the limit of large system size (\cref{sec:qke-emergence}). In \cref{sec:quantum-sim}, we describe the quantum simulations we have carried out, both on conventional hardware and on superconducting quantum processors. In this context, we set out the general formalism (\cref{sec:implementation}), develop new, highly optimized algorithms based on our previous work in Ref.~\cite{Bleau:2026iiq} (\cref{sec:bipolar-algo,sec:quantum-algo}), and we present the results of our simulation (\cref{sec:results-qke-vs-multiparticle}).
Finally, in \cref{sec:systems-with-entanglement}, we introduce two more toy systems, inspired by the actual dynamics inside a supernova, to which the preceding arguments do not apply, and in which quantum entanglement may therefore be phenomenologically relevant. Specifically, we consider a linear stream of neutrinos with overlapping wave packets in \cref{sec:neutrino-chain}, as well as systems in which multiple once-in-a-lifetime encounters lead to the build-up of entanglement in \cref{sec:once-in-a-lifetime}. We conclude in \cref{sec:conclusions}.

%=============================================================================
\section{Quantum Kinetic Equations vs.\ Multiparticle Entanglement}
\label{sec:qke-vs-multiparticle}
%=============================================================================

%-----------------------------------------------------------------------------
\subsection{The Quantum Kinetic Equations for Collective Neutrino Oscillations}
\label{sec:qke-formalism}
%-----------------------------------------------------------------------------

We begin by briefly reviewing the quantum kinetic approach to collective neutrino oscillations (sometimes called the ``mean-field approach''), based on pioneering work by Sigl and Raffelt \cite{Sigl:1993ctk}. In this formalism, the von Neumann equation describing vacuum oscillations, Mikheyev--Smirnov--Wolfenstein (MSW) matter effects \cite{Mikheyev:1985zog, Wolfenstein:1977ue}, neutrino--neutrino interactions, and neutrino--antineutrino interactions takes the form \cite{Duan:2010bg}
\begin{multline}
    i \big( \partial_t + \vec\nabla_{\vec{x}} \big) \rho_{\vec{x},\vec{p},t} \\
    = \Big[ H_0
        + \sqrt{2} G_F \int \! d^3p' \, (1 - \vec{\hat{p}} \cdot \vec{\hat{p}}')
                (\rho_{\vec{x},\vec{p'},t} - \bar\rho_{\vec{x},\vec{p'},t}), \,
        \rho_{\vec{x},\vec{p},t}
    \Big] ,
    \label{eq:von-Neumann}
\end{multline}
where $\rho_{\vec{x},\vec{p},t}$ ($\bar\rho_{\vec{x},\vec{p},t}$) is the $2 \times 2$ neutrino (antineutrino) density matrix at position $\vec{x}$, momentum $\vec{p}$, and time $t$ in the flavour basis $(\ket{\nu_e}, \ket{\nu_\mu})$. For simplicity, we work here in the two-flavour approximation, which is motivated by the fact that $\nu_\mu$ and $\nu_\tau$ behave similarly in supernovae, neutron star mergers, and the early Universe (see, however, Refs.~\cite{%
 Friedland:2010sc,       % 2f evolution unstable, 3f leads to different spectral swaps
 Dasgupta:2010cd,        % 3f spectral splits in spectral regions stable under dm31-scale 2f evolution
 Esteban-Pretel:2007ncu, % mu-tau matter potential differences modify osc., sensitivity to th23 octant
 Capozzi:2020kge,        % 3f simulations of fast conversion in background with muons
 Capozzi:2022dtr}).      % follow-up to Capozzi:2020kge; 1+1D simulations
The two terms on the left-hand side of \cref{eq:von-Neumann} describe time evolution and spatial transport, respectively.  The first term in the Hamiltonian on the right-hand side encapsulates vacuum oscillations and MSW matter effects. It is given by
\begin{align}
    H_0 \equiv U \begin{pmatrix}
                     0 & \\
                       & \frac{\delta m^2}{2E}
                 \end{pmatrix} U^\dag
        + \begin{pmatrix}
              \sqrt{2} G_F n_e \\
                               & 0
          \end{pmatrix},
\end{align}
where $E$ is the neutrino energy, $U$ the leptonic mixing matrix, $\delta m^2 \equiv m_2^2 - m_1^2$ the mass squared difference, $G_F$ the Fermi constant, and $n_e$ the net electron number density in the background matter, which controls the MSW potential. The second term on the right-hand side of \cref{eq:von-Neumann} describes neutrino self-interactions. Note in particular the angular dependence: the self-interaction term depends on the angle between the unit vectors $\vec{\hat{p}} \equiv \vec{p} / |\vec{p}|$ and $\vec{\hat{p}}' \equiv \vec{p}' / |\vec{p}'|$.

To solve \cref{eq:von-Neumann}, $\vec{x}$ and $\vec{p}$ are discretized, leading to $N$ coupled equations, one for each mode $(\vec{x}, \vec{p})$.
The main assumption that justifies working with $N$ separate $2 \times 2$ density matrices instead of a single $2^N \times 2^N$ matrix is ``molecular chaos'', that is, the assumption that the duration of each interaction is short compared to the time a neutrino takes to travel the mean free path $\lambda_\text{mfp}$ between interactions, and that any quantum correlations built up during an interaction is ``forgotten'' before the next interaction happens \cite{
    Sigl:1993ctk,    % the original paper by Sigl & Raffelt
    Johns:2023ewj,   % Neutrino many-body correlations
    Kost:2024esc}.   % Once-in-a-lifetime encounter models
\Cref{eq:von-Neumann} is not valid when a description of the system in terms of individual particles breaks down, for instance because the size of the neutrino wave packets is larger than the mean free path, or when quantum entanglement builds up over multiple interactions.

The condition for the validity of the quantum kinetic approach can be recast into the requirement that $N$ is not too small \cite{Friedland:2003dv, Friedland:2003eh}. 
The apparent contradiction to the above statement that the mean free path should be large is resolved as follows: to leave the collective forward scattering interaction strength $\propto N G_F^\text{eff}$ invariant as $N$ is varied, a change in $N$ needs to be accompanied by an opposite change in the effective coupling constant $G_F^\text{eff} \propto 1/N$. Therefore, as $\lambda_\text{mfp} \propto 1 / [(G_F^\text{eff})^2 N]$, increasing $N$ indeed \emph{increases} the mean free path.

Our goal in the following will be to explicitly demonstrate the emergence of quantum kinetic behaviour in toy systems as $N$ is increased.

%-----------------------------------------------------------------------------
\subsection{Example: Two-mode oscillations}
\label{sec:bipolar}
%-----------------------------------------------------------------------------

\begin{figure}
    \centering
    \includegraphics[width=1\linewidth]{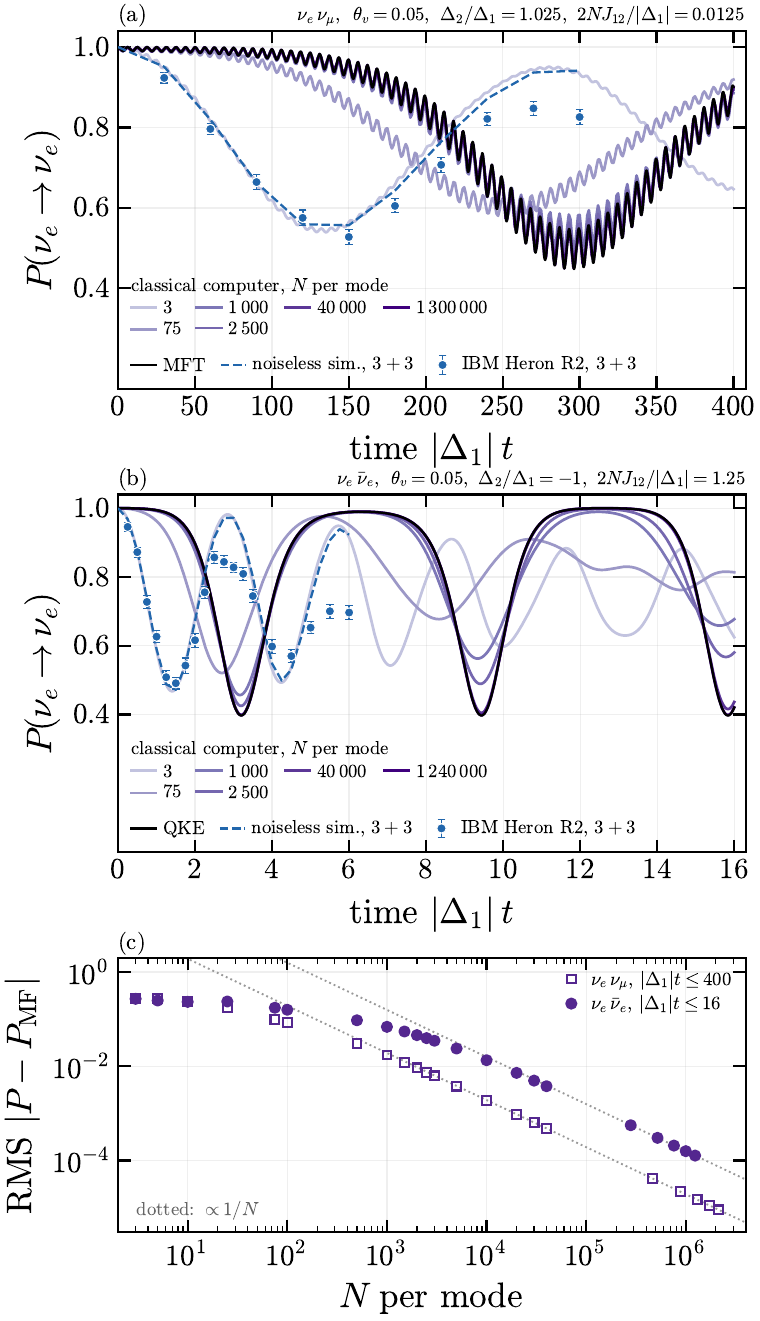}
    \caption{Comparison of the quantum kinetic mean-field treatment of collective neutrino oscillations (black curves, see \cref{sec:qke-formalism}) to the results of full multiparticle simulations (see \cref{sec:quantum-sim}) on classical hardware (purple curves) and on a superconducting transmon quantum processor (blue data points). The blue dashed line correspond to a noiseless simulation of the Trotterized quantum circuit.
    {\bf Top:} a two-mode system with initially $N$ electron neutrinos and $N$ muon neutrinos, exhibiting both slow and fast oscillations. {\bf Middle:} bipolar oscillations in a system with $N$ neutrinos and $N$ antineutrinos. {\bf Bottom:} root-mean-square residuals between the quantum kinetic mean-field results and the multiparticle simulation as a function of the number of neutrinos per mode.}
    \label{fig:bipolar}
\end{figure}

As a first example, we consider a toy system with initially $N_1$ electron neutrinos and $N_2$ muon neutrinos. This is one of the systems for which ref.~\cite{Friedland:2003dv} has demonstrated analytically the equivalence between the mean-field and multiparticle treatments. The solution of the quatum-kinetic equations, \cref{eq:von-Neumann}, is shown as a black curve in the top panel of \cref{fig:bipolar}. The other curves in this plot, showing how the full multiparticle solution converges to the quantum kinetic one in the large-$N$ limit, will be discussd in \cref{sec:quantum-sim}.

The second system we consider consists of $N_1$ neutrinos that interact with $N_2$ \emph{antineutrinos}. Flavour conversion in such a system can exhibit highly anharmonic behaviour (so-called ``bipolar'' oscillations) \cite{Hannestad:2006nj, Raffelt:2007yz, Duan:2010bg}, as shown in the middle panel of \cref{fig:bipolar}. This behaviour can be understood in the ``flavour pendulum'' picture from refs.~\cite{Hannestad:2006nj, Raffelt:2007yz}, where the neutrino and antineutrino density matrices, averaged over $\vec{x}$ and $\vec{p}$, are represented by a flavour polarization vector $\vec{P}$ in three dimensions via $P_j \equiv \tr(\rho \sigma_j)$ and $\Bar{P}_j \equiv -\tr(\bar\rho \sigma_j)$, with the Pauli matrices $\sigma_j$. In this picture, an initial state consisting of roughly equal numbers of $\nu_e$ and $\bar\nu_e$ corresponds to $\vec{P}$ and $\Bar{\vec{P}}$ pointing in opposite directions, forming a ``pendulum'' in an unstable equilibrium configuration. The small perturbation of this equilibrium configuration due to the vacuum oscillation term in the Hamiltonian leads to the sudden, anharmonic, flavour flips observed in \cref{fig:bipolar}.

In the following, we demonstrate explicitly how the quantum kinetic description of these two-mode systems emerges from the full multiparticle dynamics. We begin with analytic arguments, which we then confirm using quantum simulations in \cref{sec:quantum-sim}.

%-----------------------------------------------------------------------------
\subsection{Emergence of quantum kinetic behaviour in the large-$N$ limit}
\label{sec:qke-emergence}
%-----------------------------------------------------------------------------

Following ref.~\cite{Friedland:2003dv}, we show analytically how quantum kinetic behaviour emerges in the large-$N$ limit for bipolar oscillations between $N_1$ neutrinos and $N_2$ antineutrinos. (The analogous calculation for the $\nu_e$--$\nu_\mu$ system from the top panel of \cref{fig:bipolar} can be found in ref.~\cite{Friedland:2003dv}.) We assume that a beam of $N_1$ neutrinos are initially in a state
\begin{align}
    \nu_A &= \cos\beta\,\nu_e + \sin\beta\,\nu_{\mu},
    \label{eq:nu-A}
\intertext{while a beam of $N_2$ antineutrinos are in a state}
    \Bar{\nu}_X &= \cos\alpha\,\Bar{\nu}_e + \sin\alpha\,\Bar{\nu}_{\mu}.
    \label{eq:nu-X-bar}
\end{align}
An ensemble of neutrinos that is originally in a flavour eigenstate will eventually evolve into such a superposition due to vacuum oscillations.
We also introduce the states orthogonal to $\nu_A$ and $\Bar{\nu}_X$, which are given by
\begin{align}
    \nu_B &= -\sin\beta\,\nu_e + \cos\beta\,\nu_{\mu} ,
    &\Bar{\nu}_Y &= -\sin\alpha\,\Bar{\nu}_e + \cos\alpha\,\Bar{\nu}_{\mu}
\end{align}
and satisfy $\sprod{\nu_B}{\nu_A} = \sprod{\Bar{\nu}_Y}{\Bar{\nu}_X} = 0$. It is useful to transform $\Bar\nu_X$ into the ($A$, $B$) basis according to
\begin{align}
    \Bar\nu_X = \sprod{\Bar{\nu}_A}{\Bar{\nu}_X}\Bar\nu_A
              + \sprod{\Bar{\nu}_B}{\Bar{\nu}_X}\Bar\nu_B .
    \label{eq:nuX-as-nuAB}
\end{align}
Coherent flavour changes in this system can occur either through ordinary oscillations in matter or through coherent forward scattering. The latter can be mediated by $s$-channel interactions of the form $\nu_A \Bar{\nu}_A \to \nu_A \Bar{\nu}_A,\, \nu_B \Bar{\nu}_B$ and $t$-channel interactions like $\nu_A \Bar{\nu}_A \to \nu_A \Bar{\nu}_A$ and $\nu_A \Bar{\nu}_X \to \nu_A \Bar{\nu}_X$. $u$-channel processes (e.g.\ $\nu_A \Bar{\nu}_X \to \Bar{\nu}_X \nu_A$) cannot cause coherent interference due to the final state containing an antineutrino in the neutrino beam and vice-versa. The same is true for the additional $s$-channel processes of the form $\nu_A \Bar{\nu}_A \to \Bar{\nu}_A \nu_A,\, \Bar{\nu}_B \nu_B$.

Dropping the non-coherent processes, the neutrino interaction Hamiltonian in the $(\ket{\nu_A},\, \ket{\nu_B})$ basis is
\begin{equation}
    H_{\bar\nu\nu} = \sqrt{2} G_F n_\nu (1 - \cos\theta)
                      \begin{pmatrix}
                          -2 &        &        & \!\!\!-1\\
                             & \!\!\!-1 \\
                             &        & \!\!\!-1 \\
                          -1 &        &        & \!\!\!-2
                      \end{pmatrix}
                      {\color{gray}\ \ 
                      \begin{array}{l}
                          \ket{\nu_A\bar\nu_A} \\
                          \ket{\nu_A\bar\nu_B} \\
                          \ket{\nu_B\bar\nu_A} \\
                          \ket{\nu_B\bar\nu_B}
                      \end{array}}.
    \label{eq:H-nubarnu}
\end{equation}
Since terms proportional to the identity matrix only add a global phase under time evolution and therefore do not affect physical quantities, we can equivalently write
\begin{equation}
    H_{\bar\nu\nu} = \sqrt{2} G_F n_\nu (1 - \cos\theta)
                      \begin{pmatrix}
                          -1 &   &   & -1 \\
                             & 0 \\
                             &   & 0 \\
                          -1 &   &   & -1
                      \end{pmatrix}
                      {\color{gray}\ \ 
                      \begin{array}{l}
                          \ket{\nu_A\bar\nu_A} \\
                          \ket{\nu_A\bar\nu_B} \\
                          \ket{\nu_B\bar\nu_A} \\
                          \ket{\nu_B\bar\nu_B}
                      \end{array}}.
    \label{eq:H-nubarnu-simplified}
\end{equation}
Dropping the $-\mathbbm{1}_{4\times4}$ contribution is equivalent to neglecting the $t$-channel contributions of the form $\nu_A \Bar{\nu}_A \to \nu_A \Bar{\nu}_A$  and $\nu_A \Bar{\nu}_B \to \nu_A \Bar{\nu}_B$, which are always flavour-conserving as they are mediated by the $Z$ boson.

In one time step $\delta t$ governed by the Hamiltonian in \cref{eq:H-nubarnu-simplified}, a system consisting initially of $\nu_A$ and $\bar\nu_X$ evolves according to\footnote{Note that we consider only the time-evolution under the interaction Hamiltonian $H_{\bar\nu\nu}$ here. It is understood that time steps of the form \cref{eq:s-channel-2nu-wz} are interleaved with time steps of the form $e^{-i H_0 \delta t}$ accounting for vacuum oscillations.}
\begin{align}
    \ket{\nu_A \bar\nu_X} 
        &\xrightarrow{\delta t} \, e^{-iH_{\Bar{\nu}\nu}\delta t}
            \ket{\nu_A \bar\nu_X} \\
        &\simeq \ket{\nu_A \bar\nu_X}
           -i a \sprod{\Bar{\nu}_A}{\Bar{\nu}_X} \big(
            \ket{\nu_A \Bar{\nu}_A} + \ket{\nu_B \Bar{\nu}_B} \big) .
    \label{eq:s-channel-2nu-wz}
\end{align}
Here, we have used the fact that $H_{\bar\nu\nu}$ acts non-trivially only on the $\ket{\nu_A\bar\nu_A}$ component of the initial state, while leaving the $\ket{\nu_A\bar\nu_B}$ component invariant. We have moreover defined $a \equiv -\delta t \sqrt{2} G_F n_{\nu} (1 - \cos\theta)$.

We now generalize the two-particle interaction from \cref{eq:s-channel-2nu-wz} to the full $(N_1+N_2)$-particle system. To simplify the notation, we write $\ket{\nu_A} \equiv \ket{A}$ in the following, and analogously for $\nu_B$, $\nu_X$, $\nu_Y$. The evolution of the full system during an infinitesimal time step is given by
\begin{align}
    \ket{AAA...}\ket{\Bar{X}\Bar{X}\Bar{X}...} \xrightarrow{\delta t}
    \ket{F} \equiv \ket{AAA...}\ket{\Bar{X}\Bar{X}\Bar{X}...} - i a \ket{F_1},
    \label{eq:time-step-full-system}
\end{align}
where
\begin{widetext}
\begin{align}
    \ket{F_1} = &\Big[ \big(\ket{AAA...} + \ket{BAA...}\big)
                     + \big(\ket{AAA...} + \ket{ABA...}\big) + ... \Big] \notag\\
    &\qquad\qquad
    \times \sprod{\Bar{A}}{\Bar{X}} \Big[
        \big(\ket{\Bar{A}\Bar{X}\Bar{X}...} + \ket{\Bar{B}\Bar{X}\Bar{X}...}\big) + \big(\ket{\Bar{X}\Bar{A}\Bar{X}...} + \ket{\Bar{X}\Bar{B}\Bar{X}...}\big) + ...
    \Big] .
    \label{eq:s-channel-2nu-xy}
\end{align}
This expression sums the $s$-channel interaction from \cref{eq:s-channel-2nu-wz} over all $N_1 N_2$ neutrino--antineutrino combinations. We now express $\ket{\Bar{A}}$ and $\ket{\Bar{B}}$ in the $(\ket{\Bar{X}}, \ket{\Bar{Y}})$ basis:
\begin{align}
    \ket{F_1} &= N_1 \sprod{\Bar{A}}{\Bar{X}} \ket{AAA...} \Big[
        \Big(\!\sprod{\Bar{X}}{\Bar{A}} \ket{\Bar{X}\Bar{X}\Bar{X}...}
             + \sprod{\Bar{Y}}{\Bar{A}} \ket{\Bar{Y}\Bar{X}\Bar{X}...} \Big)
      + \Big(\!\sprod{\Bar{X}}{\Bar{A}} \ket{\Bar{X}\Bar{X}\Bar{X}...}
             + \sprod{\Bar{Y}}{\Bar{A}} \ket{\Bar{X}\Bar{Y}\Bar{X}...} \Big) + ...
      \Big]                                         \nonumber \\
    &\quad 
     + \sprod{\Bar{A}}{\Bar{X}} \Big[ \ket{BAA...} + \ket{ABA...} + ... \Big] \!
       \Big[
        \Big(\!\sprod{\Bar{X}}{\Bar{B}} \ket{\Bar{X}\Bar{X}\Bar{X}...}
             + \sprod{\Bar{Y}}{\Bar{B}} \ket{\Bar{Y}\Bar{X}\Bar{X}...} \Big)
                                                    \nonumber \\
    &\hspace{5.7cm}
      + \Big(\!\sprod{\Bar{X}}{\Bar{B}} \ket{\Bar{X}\Bar{X}\Bar{X}...}
             + \sprod{\Bar{Y}}{\Bar{B}} \ket{\Bar{X}\Bar{Y}\Bar{X}...} \Big) + ...
       \Big] \label{eq:F1_1} \\
    &= \sprod{\Bar{A}}{\Bar{X}} \Big[
          N_1 N_2 \sprod{\Bar{X}}{\Bar{A}} \ket{AAA...} \ket{\Bar{X}\Bar{X}\Bar{X}...}
                                                    \nonumber \\
    &\quad
        + N_1 \sprod{\Bar{Y}}{\Bar{A}} \ket{AAA...}
          \Big( \ket{\Bar{Y}\Bar{X}\Bar{X}...}
              + \ket{\Bar{X}\Bar{Y}\Bar{X}...} + ... \Big)
        + N_2 \sprod{\Bar{X}}{\Bar{B}}
          \Big( \ket{BAA...} + \ket{ABA...} + ... \Big)
          \ket{\Bar{X}\Bar{X}\Bar{X}...}            \nonumber \\
    &\quad
        + \sprod{\Bar{Y}}{\Bar{B}} \Big( \ket{BAA...} + \ket{ABA...} + ... \Big)
          \Big( \ket{\Bar{Y}\Bar{X}\Bar{X}...} + \ket{\Bar{X}\Bar{Y}\Bar{X}...} + ...\Big)
       \Big],
    \label{eq:F1_2}
\end{align}
When computing flavour transition probabilities, we are ultimately interested in expectation values of the form $\bra{F} \hat{\alpha} \ket{F}$, where $\hat{\alpha}$ is the operator that counts the number of $\ket{\nu_\alpha}$ in the neutrino ensemble ($\alpha = e, \mu$). It is given by 
\begin{align}
    \hat{\alpha} \equiv \sum_{j=1}^{N_1} \ket{\alpha_j} \bra{\alpha_j},
\end{align}
noting that $\ket{\alpha_j} \bra{\alpha_j}$ projects the $j$-th neutrino onto the flavour eigenstate $\ket{\nu_\alpha}$. There is a similar operator $\hat{\Bar\alpha}$ for antineutrinos. Using \cref{eq:time-step-full-system,eq:F1_2}, we find
\begin{align}
    \bra{F} \hat{\alpha} \ket{F} = N_1 \big|\! \sprod{\alpha}{A} \!\big|^2 + a^2 \bra{F_1} \hat{\alpha} \ket{F_1},
\end{align}
where
\begin{align}
    &\bra{F_1} \hat{\alpha} \ket{F_1} = \big|\! \sprod{\Bar{A}}{\Bar{X}} \!\big|^2 \times
            \Big[
               N_1^3 N_2^2 \, \big|\! \sprod{\Bar{A}}{\Bar{X}} \!\big|^2 \,
                              \big|\! \sprod{\alpha}{A} \!\big|^2
             + 2 \Re\Big(
               N_1^2 N_2^2 \sprod{\Bar{X}}{\Bar{A}} \sprod{\Bar{B}}{\Bar{X}}
                           \sprod{B}{\alpha} \sprod{\alpha}{A} \Big)
                                                    \nonumber\\
    &\ \
             + N_1^3 N_2 \big|\! \sprod{\Bar{Y}}{\Bar{A}} \!\big|^2 \,
                         \big|\! \sprod{\alpha}{A} \!\big|^2
             + 2 \Re\Big(\!
               N_1^2 N_2 \sprod{\Bar{Y}}{\Bar{A}} \sprod{\Bar{B}}{\Bar{Y}}
                         \sprod{B}{\alpha} \sprod{\alpha}{A} \!\!\Big)
             + N_1 N_2^2 \big|\! \sprod{\Bar{X}}{\Bar{B}} \!\big|^2 \, \Big(
                             \big|\! \sprod{B}{\alpha} \!\big|^2
                                   + (N_1-1) \big|\! \sprod{A}{\alpha} \!\big|^2
                         \Big)                      \nonumber\\
    &\ \
             + N_1 N_2  \big|\! \sprod{\Bar{B}}{\Bar{Y}} \!\big|^2 \, \Big(
                            \big|\! \sprod{B}{\alpha} \!\big|^2
                                + (N_1-1) \big|\! \sprod{A}{\alpha} \!\big|^2
                        \Big)
    \Big].
    \label{eq:ev-alpha}
\end{align}
\end{widetext}
Dropping terms that are not coherently enhanced -- that is, terms that are not proportional to at least four powers of $N_1$ or $N_2$ -- we see that the last three terms in \cref{eq:ev-alpha} are negligible and can be removed. Therefore, for practical purposes, the last term in \cref{eq:F1_2} can also be dropped, which significantly simplifies $\ket{F_1}$. Notably, we can now express the final state after an infinitesimal time step as a product state of the form
\begin{align}
    \ket{F} = \ket{A'A'A'...} \ket{\Bar{X}'\Bar{X}'\Bar{X}'...}
\end{align}
up to first order in $a$, where
\begin{align}
    \ket{A'} &= \ket{A} - i a N_2 \Big(
        \tfrac{1}{2} \big|\! \sprod{\Bar{A}}{\Bar{X}} \!\big|^2 \, \ket{A}
      + \sprod{\Bar{A}}{\Bar{X}} \sprod{\Bar{X}}{\Bar{B}} \, \ket{B} \Big), \\
    \ket{\Bar{X}'} &= \ket{\Bar{X}} - i a N_1 \Big(
        \tfrac{1}{2} \big|\! \sprod{\Bar{A}}{\Bar{X}} \!\big|^2 \, \ket{\Bar{X}}
      + \sprod{\Bar{A}}{\Bar{X}} \sprod{\Bar{Y}}{\Bar{A}} \, \ket{\Bar{Y}} \Big).
\end{align}
This shows that, in the limit of large $N_1$ and $N_2$, the system can be written as a product state \emph{without entanglement} after a single time step. Relabelling $\ket{A'} \to \ket{A}$ and $\ket{\Bar{X}'} \to \ket{\Bar{X}}$ for the next time step, we see that factorization continues to hold after an arbitrary number of time steps. Consequently, the system never develops significant entanglement, which justifies a description using quantum kinetic equations.

\vspace{1em}
One may wonder whether the picture changes if entanglement is already present in the initial state. (For the moment, we leave aside the question of how such entanglement could have been created in a realistic astrophysical or cosmological environment.) As an example, we introduce an entangled state of the form
\begin{align}
    \ket{AAA...} \rightarrow \ket{AAA...} + \epsilon \ket{BBB...},
\end{align}
where $\epsilon$ parameterizes the amount of entanglement. The interaction between the neutrino and antineutrino beams becomes
\begin{widetext}
\begin{align}
    \big( \ket{AAA...} + \epsilon \ket{BBB...} \big) \ket{\Bar{X}\Bar{X}\Bar{X}...} \xrightarrow{\delta t}
    \ket{F} \equiv
    \big( \ket{AAA...} + \epsilon \ket{BBB...}) \ket{\Bar{X}\Bar{X}\Bar{X}...}
  - i a \big( \ket{F_1} + \epsilon \ket{F_1'} \big),
\end{align}
where $\ket{F_1}$ is given by \cref{eq:F1_2} and $\ket{F_1'}$ reads, in analogy to \cref{eq:F1_2},
\begin{align}
    \ket{F_1'}
    &= \sprod{\Bar{B}}{\Bar{X}} \Big[
          N_1 N_2 \sprod{\Bar{X}}{\Bar{B}} \ket{BBB...} \ket{\Bar{X}\Bar{X}\Bar{X}...}
                                                    \nonumber \\
    &\quad
        + N_1 \sprod{\Bar{Y}}{\Bar{B}} \ket{BBB...}
          \Big( \ket{\Bar{Y}\Bar{X}\Bar{X}...}
              + \ket{\Bar{X}\Bar{Y}\Bar{X}...} + ... \Big)
        + N_2 \sprod{\Bar{X}}{\Bar{A}}
          \Big( \ket{ABB...} + \ket{BAB...} + ... \Big)
          \ket{\Bar{X}\Bar{X}\Bar{X}...}            \nonumber \\
    &\quad
        + \sprod{\Bar{Y}}{\Bar{A}} \Big( \ket{ABB...} + \ket{BAB...} + ... \Big)
          \Big( \ket{\Bar{Y}\Bar{X}\Bar{X}...} + \ket{\Bar{X}\Bar{Y}\Bar{X}...} + ...\Big)
       \Big].
    \label{eq:F1_2_entangled}
\end{align}
\end{widetext}
As in \cref{eq:F1_2}, the last term is not amplified coherently and can be dropped. We can then once more express the final state as a sum of product states
\begin{align}
    \ket{F} = \ket{A'A'A'...}\ket{\Bar{X}'\Bar{X}'\Bar{X}'...} + \epsilon \ket{B'B'B'...}\ket{\Bar{X}''\Bar{X}''\Bar{X}''...},
\end{align}
where 
\begin{align}
    \ket{A'} &= \ket{A} - i a N_2 \Big(
        \tfrac{1}{2} \big|\! \sprod{\Bar{A}}{\Bar{X}} \!\big|^2 \, \ket{A}
      + \sprod{\Bar{A}}{\Bar{X}} \! \sprod{\Bar{X}}{\Bar{B}} \, \ket{B} \Big), \\
    \ket{B'} &= \ket{B} - i a N_2 \Big(
        \tfrac{1}{2} \big|\! \sprod{\Bar{B}}{\Bar{X}} \!\big|^2 \, \ket{B}
      + \sprod{\Bar{B}}{\Bar{X}} \! \sprod{\Bar{X}}{\Bar{A}} \, \ket{A} \Big), \\
    \ket{\Bar{X}'} &= \ket{\Bar{X}} - i a N_1 \Big(
        \tfrac{1}{2} \big|\! \sprod{\Bar{A}}{\Bar{X}} \!\big|^2 \, \ket{\Bar{X}}
      + \sprod{\Bar{A}}{\Bar{X}} \! \sprod{\Bar{Y}}{\Bar{A}} \, \ket{\Bar{Y}} \Big), \\
    \ket{\Bar{X}''} &= \ket{\Bar{X}} - i a N_1 \Big(
        \tfrac{1}{2} \big|\! \sprod{\Bar{B}}{\Bar{X}} \!\big|^2 \, \ket{\Bar{X}}
      + \sprod{\Bar{B}}{\Bar{X}} \! \sprod{\Bar{Y}}{\Bar{B}} \, \ket{\Bar{Y}} \Big).
\end{align}
Therefore, despite the initial state being entangled, no additional entanglement builds up over time. In the bipolar system, the quantum kinetic approach is hence justified in the large $N_1$ and $N_2$ limit even in the case where entanglement is initially present. We expect that this conclusion generalizes to more complex systems.

%=============================================================================
\section{Quantum Simulation of Multiparticle Dynamics}
\label{sec:quantum-sim}
%=============================================================================

Having demonstrated analytically the emergence of the mean-field limit at large $N_1$ and $N_2$ in simple toy systems, we now confirm this result using full multiparticle simulations with entanglement on both classical and quantum hardware.

%-----------------------------------------------------------------------------
\subsection{Quantum Circuits for General Systems}
\label{sec:implementation}
%-----------------------------------------------------------------------------

In the simplest quantum circuits for collective neutrino oscillations in the two-flavour approximation, the flavour eigenstates $\ket{\nu_e}$ and $\ket{\nu_\mu}$ of each neutrino are mapped onto qubit states according to $\ket{\nu_e} \leftrightarrow \ket{0}$, $\ket{\nu_\mu} \leftrightarrow \ket{1}$. The Hamiltonian of the system is
\begin{align}
    \hat{H}_q \equiv \hat{H}_{q,0} + \hat{H}_{q,\nu\nu}.
    \label{eq:H-q}
\end{align}
The first term, describing vacuum oscillations and MSW interactions with ordinary matter, has the form
\begin{align}
    \hat{H}_{q,0} \equiv \frac{1}{2} \sum_p \vec{b}_p \cdot \hat{\vec\sigma}_p,
    \label{eq:Hq0}
\end{align}
where $\hat{\vec\sigma}_p \equiv (\hat\sigma_p^x, \hat\sigma_p^y, \hat\sigma_p^z)$ is the vector of Pauli matrices acting on the $p$-th qubit and $\vec{b}_p \equiv \Delta_p (\sin 2\theta, 0, -\cos 2\theta)$, where $\Delta_p = \delta m^2 / (2 E_p)$ is the oscillation frequency of the $p$-th neutrino. $\theta$ is the mixing angle and the direction of $\vec{b}_p$ in vacuum is only dependent on the mixing between lighter and heavier neutrino eigenstates. Both $\delta m^2$ and $\theta$ are understood to be taken at their effective values in matter. The second term, describing self-interactions, is
\begin{align}
    \hat{H}_{q,\nu\nu} \equiv \sum_{p<q} J_{pq} \hat{\vec\sigma}_p \otimes \hat{\vec\sigma}_q,
    \label{eq:Hqnunu}
\end{align}
where $J_{pq} \equiv (G_F n_\nu / \sqrt{2}) (1 - \cos\alpha_{pq})$ is the self-interaction strength and $\alpha_{pq}$ is the angle between the momentum vectors of modes $p$ and $q$.
In this paper, we will not concern ourselves with angular effects and simply treat $J_{pq}$ as a free coupling constant.
For neutrino--antineutrino interactions, the self-interaction term is replaced by
\begin{align}
    \hat{H}_{q,\nu\bar\nu} \equiv
        \sum_{p<q} J_{pq} \big( - \hat\sigma_p^x \otimes \hat\sigma_q^x
                                + \hat\sigma_p^y \otimes \hat\sigma_q^y
                                - \hat\sigma_p^z \otimes \hat\sigma_q^z \big).
    \label{eq:H-q-nubarnu}
\end{align}
Since $H_{q,\nu\bar\nu} = (\mathbbm{1}_2 \otimes \hat\sigma_2) \cdot H_{q,\nu\nu} \cdot (\mathbbm{1}_2 \otimes \hat\sigma_2)^\dag$, it is possible to unify the description of $\nu\nu$, $\nu\bar\nu$, and $\bar\nu\bar\nu$ interactions by changing the mapping between flavour eigenstates and qubit states for antineutrinos to $\ket{\bar\nu_e} \leftrightarrow \ket{1}$, $\ket{\bar\nu_\mu} \leftrightarrow -\ket{0}$ and replacing $\vec{b}_p \to -\vec{b}_p$ for antineutrinos. After these changes, \smash{$\hat{H}_{q,\nu\bar\nu} = \hat{H}_{q,\nu\nu}$}.

The $N$-qubit quantum state $\ket{\Psi} \equiv \ket{\psi_i} \otimes \ldots \otimes \ket{\psi_n}$ is evolved in time using a second-order Suzuki--Trotter decomposition:
\begin{align}
    \ket{\Psi(t)} &= \prod_{j=1}^{n_\text{steps}} \prod_{p<q} e^{-i \hat{h}_{pq} t/n_\text{steps}} \, \ket{\Psi(0)}
    \label{eq:Trotter}
\intertext{with}
    \hat{h}_{pq} &\equiv \frac{1}{2} \frac{\vec{b}_p \cdot \hat{\vec\sigma}_p + \vec{b}_q \cdot \hat{\vec\sigma}_q}{n_\text{steps}-1}
                 + J_{pq} \hat{\vec\sigma}_p \otimes \hat{\vec\sigma}_q,
    \label{eq:hpq}
\end{align}
where $n_\text{steps}$ is the number of time steps.

Since the Hamiltonian in Eq. \ref{eq:H-q} is invariant under permutations of neutrinos that share the same 4-momentum, i.e.\ the same $\vec{b}_p \equiv \vec{b}$ and $J_{pq} \equiv J$, we can use the correspondence between qubits and spin-$\frac{1}{2}$ particles to rewrite it as \cite{Bleau:2026iiq}
\begin{align}
    \hat{H}_s \equiv \vec{b} \cdot \hat{\vec{S}}_\text{tot}
             + 2 \, J \, \hat{\vec{S}}_\text{tot}^2,
    \label{eq:Hs-1-mode}
\end{align}
with $\hat{\vec{S}}_\text{tot}$ the total spin vector. If the system is initially in an eigenstate of $\hat{\vec{S}}^{2}_\text{tot}$ and $\hat{S}_{\text{tot},z}$, a so-called Dicke state, the second term can be dropped as $\hat{\vec{S}}^{2}_\text{tot}$ is a conserved quantity. In order to describe  multiple neutrino modes with different energies and propagation directions, and/or a system with mixtures of neutrinos and antineutrinos, \cref{eq:Hs-1-mode} can be generalized to 
\begin{align}
    \hat{H}_s \equiv \sum_i \vec{b}_i \cdot \hat{\vec{S}}_i
             + 2 \sum_i J_i \hat{\vec{S}}_i^2 
             + 4 \sum_{i<j} J_{ij} \hat{\vec{S}}_i \cdot \hat{\vec{S}}_j,
    \label{eq:Hs-n-modes}
\end{align}
where $\vec{b}_i$ parameterizes vacuum oscillations and MSW matter effects with ordinary matter of the $i$-th mode, $J_i$ is the self-interaction strength among neutrinos in mode $i$, and $J_{ij}$ is the interaction strength between modes $i$ and $j$. If the system starts out in an eigenstate of the $\hat{\vec{S}}_i^2$, the self-interaction terms can be dropped.

To encode Dicke states in quantum registers, we split the mode comprised of $N_i$ neutrinos into multiple new modes which are eigenstates of the corresponding $\vec{S}_i^2$ with quantum numbers $S_i = N_i/2$ and $m_i = \pm S_i$. The Hilbert space of the $i$-th mode is restricted by permutation symmetry to a subspace spanned by $N_i+1$ states $\ket{S_i, m_i}$ with $m_i = -S_i, -S_i+1, \ldots, +S_i$. This state can be encoded as a binary number in a register consisting of $\lceil \log_2(N_i+1) \rceil$ qubits via the computational-basis mapping
\begin{align}
    \ket{j}_{\text{comp}} \;\longleftrightarrow\;
    \ket{S,\; m = j - S},
    \label{eq:log_encoding}
\end{align}
with $j \in \{0, 1, \ldots, N\}$, and prepared from a state in which all qubits are in the $\ket{0}$ state by applying a series of Pauli $X$ gates. To evolve the system with time according to the Hamiltonian in Eq. \ref{eq:Hs-n-modes}, we also generalize the $RX$, $RZ$, $R_{XX}$, $R_{YY}$, and $R_{ZZ}$ gates to Dicke states.

The Dicke approach, which exploits the symmetry of the Hamiltonian and relates it to $su(2)$ algebra, is computationally advantageous, since the complexity of the Hilbert space on which \cref{eq:Hs-n-modes} acts grows exponentially with the number of \emph{modes}, while the complexity of the original Hamiltonian, \cref{eq:H-q}, is exponential in the number of neutrinos. More details about conventional circuits for collective neutrino oscillations, Dicke states, and the encoding of Dicke states in quantum registers are given in ref.~\cite{Bleau:2026iiq}.

%-----------------------------------------------------------------------------
\subsection{Optimized Multiparticle Quantum Simulation of Bipolar Oscillations on Classical Hardware}
\label{sec:bipolar-algo}
%-----------------------------------------------------------------------------

The two-mode oscillation scenarios introduced in \cref{sec:qke-vs-multiparticle} feature a Hamiltonian structure that allows us to evolve the full multiparticle system on classical hardware even for $N > 10^6$. (Quantum-kinetic mean-field behaviour begins to emerge at $N \sim 100$--$1\,000$ as we will see shortly.) In the following, we outline our algorithm to achieve this.

Consider for definiteness a system with $N$ neutrinos and $N$ antineutrinos, described according to the conventions outlined above as two collective spin vectors $\vec{S}_1$, $\vec{S}_2$ with opposite vacuum oscillation terms, i.e., $\Delta_{1,2} = \mp \Delta$ for $\Delta \equiv |\delta m^2|/(2E)$. (For $N+N$ neutrinos and no antineutrinos, this would become $\Delta_1 = \Delta_2 \equiv \Delta$; the derivation remains otherwise unchanged.) From \cref{eq:Hs-n-modes}, the Hamiltonian of the bipolar system is
\begin{align}
    \hat{H} \equiv -\vec{b} \cdot (\hat{\vec{S}}_1 - \hat{\vec{S}}_2)
                 + 4 J_{12} \hat{\vec{S}}_1 \cdot \hat{\vec{S}}_2 ,
    \label{eq:Hs-2-modes}
\end{align}
with $\vec{b} \equiv |\Delta|\hat{\vec{b}}$ and $\hat{\vec{b}} \equiv (\sin 2\theta, 0, -\cos 2\theta)$.
We have again dropped the self-interaction terms, assuming the system to be in an eigenstate of $\hat{\vec{S}}_1^2$ and $\hat{\vec{S}}_2^2$. These terms commute with $\vec{S}_{\rm tot}^2$, so assuming that the initial state is an eigenstate of \smash{$\hat{\vec{S}}_{\rm tot}^2$} with eigenvalue $J (J+1)$, the system will never leave the subspace corresponding to that eigenvalue.

In the limit of large self-coupling, where vacuum oscillations can be neglected ($\vec{b} = 0$), the first (vacuum) term in \cref{eq:Hs-2-modes} is negligible. In this case, the total flavour isospin along the $z$ direction, $S_{1,z} + S_{2,z}$ is conserved in the bipolar scenario~\cite{Xiong:2021evk, Bleau:2026iiq}.

When the vacuum mixing angle $\theta \neq 0$, it is no longer the flavour isospin along the $z$ axis that is conserved, but instead the flavour isospin along a tilted axis. Namely, $[\hat{H}, \hat{\vec{b}} \cdot (\hat{\vec{S}}_1 + \hat{\vec{S}}_2)] = 0$. Defining
\begin{align}
    \hat{S}_{i,z}' \equiv \hat{\vec{b}} \cdot \hat{\vec{S}}_i, \quad i = 1,2,
    \label{eq:S-i-z-prime}
\end{align}
the conserved quantity is given by the operator
\begin{align}
    \hat{M}' \equiv \hat{S}_{1,z}' + \hat{S}_{2,z}' .
    \label{eq:M-prime}
\end{align}
The new quantization axis corresponds to working in the mass basis rather than the flavor basis. In the flavour basis, an $s=1/2$ spin oriented along the positive $z$-axis ($m = +\frac12$) corresponds to a $\nu_e$, while in the mass basis $m= +\frac12$ indicates the $\nu_2$ mass eigenstate.
(We assume an inverted mass ordering here.)

In the unprimed basis, the initial state of the bipolar system with $N$ $\nu_e$ and $N$ $\bar\nu_e$ corresponds to the neutrinos being in the $\ket{N/2, N/2}$ eigenstate of $\hat{\vec{S}}_1$, and the antineutrinos in the $\ket{N/2, -N/2}$ eigenstate of $\hat{\vec{S}}_2$.
In other words, each mode is in the Dicke state of maximal weight along the flavour axis.
After a rotation into the primed basis, it remains a spin coherent state~\cite{PhysRevA.6.2211} with the highest-weight state rotated by the angle $\pi - 2\theta$ between the unprimed and primed axes.
It remains a product state of $N$ identical spins, but it is not an eigenstate of $\hat{M}'$. In the primed basis, a maximal-weight Dicke state $\ket{J,J}$ expands to \cite{PhysRevA.105.032620}
\begin{align}
    \ket{J,J} &= \sum_{M'=-J}^{J} d^J_{M'J}(\pi-2\theta) \ket{J,M'}' ,
    \label{eq:spin-coherent-expansion}
\intertext{with}
    \big| d^J_{M'S}(\pi-2\theta) \big|^2
      &= \binom{2J}{J+M'} \sin^{2(J+M')}\!\theta \, \cos^{2(J-M')}\!\theta ,
    \label{eq:d-J-Mp-S}
\end{align}
which is a binomial distribution. In the absence of vacuum oscillations ($\theta=0$), the old and new $z$-axes would be identical, and only the term $M' = M = J$ in \cref{eq:spin-coherent-expansion} would contribute.

As shown in \cref{sec:H-bipolar-explicit}, the Hamiltonian in the primed basis has a tridiagonal structure, with the diagonal elements given by
\begin{align}
    \bra{JM'} \hat{H} \ket{JM'}
    &= 2J_{12}\left[J(J+1)-2s(s+1)\right], 
    \label{eq:H-bipolar-diag-final}
\end{align}
and the off-diagonal ones by
\begin{multline}
    \bra{J-1,M'} \hat{H} \ket{JM'} = \bra{JM'} \hat{H} \ket{J-1,M'} \\[6pt]
    = -|\Delta| \, \sqrt{
      \frac{(J^2-M^2)\left[(2s+1)^2-J^2\right]}
           {4J^2-1} }.
    \label{eq:H-bipolar-offdiag-final}
\end{multline}
On classical hardware, a sizable computational optimization is gained from evolving only the most relevant eigenstates of $\hat{M}'$ in time and dropping those which have only a small overlap with the initial state. We now briefly explain this procedure, with more details deferred to \cref{sec:adaptive-classical}.

The first step of the algorithm involves transforming the initial state from the flavour basis into the mass basis following \cref{eq:spin-coherent-expansion}.  Then, if vacuum oscillations are subdominant compared to a strong interaction term, a limited set of $\hat M'$ subspaces carries most of the initial state amplitude, since the initial state is close to the highest-weight state of $\hat M'$. Therefore, in performing the transformation in \cref{eq:spin-coherent-expansion}, we restrict the sum over $M'$ to a contiguous set of $M'$ sectors capturing all but a preset tolerance $\epsilon_M = 10^{-12}$ of the initial wave function norm. Only these are evolved in time using the Chebyshev expansion method~\cite{1984JChPh..81.3967T}, under the Hamiltonian defined by \cref{eq:H-bipolar-diag-final,eq:H-bipolar-offdiag-final}. The components of the initial state that fall outside this $M'$ range are dropped. The smaller the vacuum term, the fewer $\hat M'$ subspaces are needed to simulate the system.

A further optimization is possible during time-evolution by restricting the simulation in each $M'$ sector to only the most relevant values $J$ states. Further details on this are given in \cref{sec:adaptive-classical}.

Simulating two-mode systems with $> 10^6$ neutrinos in each mode then becomes possible, and the resulting oscillation probabilities are shown in \cref{fig:bipolar}. We will discuss these results in more detail in \cref{sec:results-qke-vs-multiparticle} below. A total of 1\,856 CPU hours was used to produce the results shown in the top panel of~\cref{fig:bipolar}. 9\,548 CPU hours were used to produce the result in the middle panel. The neutrino--antineutrino case took noticeable longer to compute since the stronger interaction in that scenario required finer time steps. Additionally, a wider range of $J$ states needs to be tracked in this case: 94\% of all $J$ states are ``active'' in the $\nu$--$\bar\nu$ system, compared to only 57\% for the $\nu_e$-$\nu_\mu$ system. The bottom panel of \cref{fig:bipolar} required 12\,239 CPU hours for the $\nu_e$--$\nu_\mu$ curves and 19\,068 CPU hours for the $\nu$--$\bar\nu$ curve..

%-----------------------------------------------------------------------------
\subsection{Quantum Algorithms with Efficient Encoding}
\label{sec:quantum-algo}
%-----------------------------------------------------------------------------

On quantum hardware, a truncation scheme like the one described in \cref{sec:bipolar-algo} would be difficult to implement, but it is also unnecessary. Instead, we need an optimization strategy that minimizes the number of (noisy) 2-qubit entangling gates and keeps the number of required qubits small. To achieve the second goal, we use the encoding from~\cref{eq:log_encoding} again. For a two-mode system with $N+N$ particles, this leads to two quantum registers consisting of $n$ physical qubits each and providing a computational basis with dimension $D = 2^n$. Each register can represent the flavor state of $N \equiv D-1$ physical neutrinos, encoded as a binary number, where the number $N$ corresponds to a state with all electron neutrinos (all muon antineutrinos), and the number $0$ corresponds to all muon neutrinos (all electron antineutrinos). Expressed in terms of Dicke states, the $N$ neutrinos are represented by a spin vector of length $j(j+1)$, with $j = N/2$, and register~$i$ encodes the integer
\begin{align}
    k_i \equiv j + m_i \in (0, 1, \ldots, N)
    \label{eq:register-integer}
\end{align}
where $m_i = -j, \ldots, j$ is the magnetic quantum number of mode~$i$.
For both toy systems from \cref{sec:qke-vs-multiparticle}, register~1 is initialized to $k_1=N$, corresponding to $N$ electron neutrinos, while the initial state of register~2 is $k_2=0$ corresponding to $N$ muon neutrinos or $N$ electron antineutrinos. We write this initial state as $\ket{N, 0}$ and a general basis state as $\ket{k_1 k_2}$.
We use the convention that the $p$-th qubit of a register carries the $p$-th bit of the number $k_i$, least significant digit first.
In the mass basis (i.e., the primed frame from \cref{sec:bipolar-algo}), the total flavour quantum number $m_1 + m_2$, and hence $k_1+k_2$ is conserved.

\begin{figure}
    \centering
    \includegraphics[width=1\linewidth]{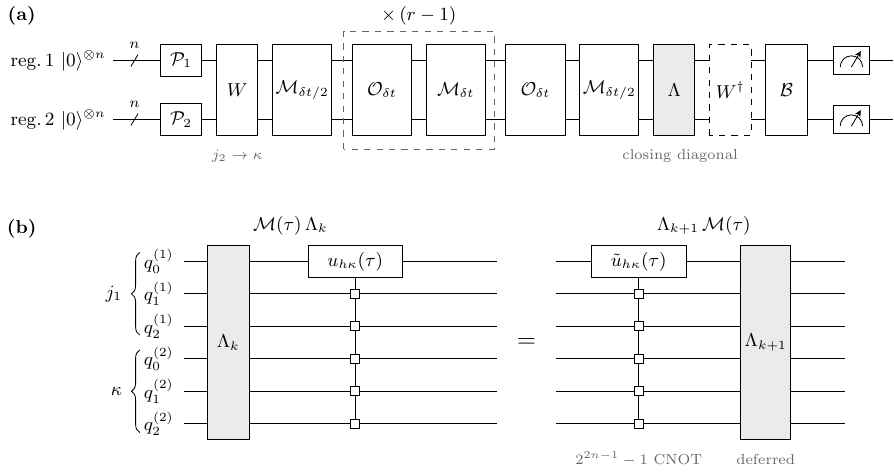}
    \caption{(a) Schematic circuit diagram of the quantum algorithm introduced in \cref{sec:quantum-algo}. $\mathcal{P}_{1,2}$ performs the initial state preparation. $W$ is a Draper adder for transforming the second register according to \cref{eq:W-transform}. $\tilde{M}$ denotes the time evolution under $\hat{H}_M$ (even bonds) for a single Trotter step, with a diagonal matrix $\Lambda_k$ removed. Similarly, $\tilde{O}$ denotes the time evolution under $\hat{H}_O$ with its diagonals removed in the same way. $\tilde{O}$ includes the decrement and increment of register~1. The $\tilde{O}_{\delta t}$--$\tilde{M}_{\delta t}$ sequence is repeated $r-1$ times, where $r$ is the number of Trotter steps. $\Lambda$ is the single deferred diagonal operator. $\mathcal{B}$ implements the measurement basis and performs single qubit rotations to measure in the $X$, $Y$, or $Z$ basis. The shown circuit applies a second order Trotter decomposition and subsequently merges adjacent half-step even blocks, leaving only the first and last half-step blocks.
    (b) Structure of the sub-circuit for a single Trotter step on the even bonds for $N=7$ with three qubits in each register. A pending diagonal $\Lambda_k$ followed by the exact multiplexer $\mathcal{M}$ of~\cref{eq:even-bond-circuit} is equivalent to a modified multiplexer $\tilde{\mathcal{M}}$ followed by a new pending diagonal $\Lambda_{k+1}$. This replacement is applied recursively across subsequent layers to defer the diagonal operators until the final closing diagonal.
    }
    \label{fig:circuit}
\end{figure}

Based on ref.~\cite{Bleau:2026iiq}, we introduce in the following an efficient algorithm for evolving the bipolar system in time using the above encoding. While tailored towards oscillations in two-mode scenarios, the algorithm remains generalizable to systems with more than two modes. A schematic diagram of the circuit is presented in~\cref{fig:circuit}.

\paragraph{Sum register.}
The fact that $k_1 + k_2$ is conserved can be exploited by applying the transformation \cite{Stryker:2021asy}
\begin{align}
    \ket{k_1 k_2} \to W \ket{k_1 k_2} \equiv \ket{k_1, \kappa}
    \label{eq:W-transform}
\intertext{with}
    \kappa \equiv (k_1 + k_2) \bmod D .
\end{align}
Afterwards, time-evolution affects only the first register. The inverse transformation, applied at the end of the circuit, is given by $k_2 = (\kappa - k_1) \bmod D$. It is exact since $k_1, k_2 < D$. The transformation can be implemented using the QFT-based adder of ref.~\cite{Draper:2000vuj}, which applies a quantum Fourier transform to one register, a sequence of controlled phase rotations conditioned on the bits of the other register, and an inverse quantum Fourier transform. It does not require ancillary qubits.

\paragraph{Hamiltonian.}
Unlike in \cref{sec:bipolar-algo}, where we worked with eigenstates of the total angular momentum operator $\hat{\vec{S}}_{\rm tot}$, it is more convenient here to retain the individual quantum numbers under $\hat{\vec{S}}_1$ and $\hat{\vec{S}}_2$. We write the Hamiltonian of the bipolar system, \cref{eq:Hs-2-modes}, as
\begin{align}
    \hat{H} \equiv -\Delta (\hat{S}_{1,z} - \hat{S}_{2,z})
                 + 2 J_{12} \big[ 2 \hat{S}_{1,z} \hat{S}_{2,z} 
                                + \hat{S}_1^+ \hat{S}_2^- + \hat{S}_1^- \hat{S}_2^+
                            \big]
    \label{eq:Hs-2-modes-ladder}    
\end{align}
with the ladder operators $\hat{S}_k^\pm \equiv \hat{S}_{k,x} \pm i \hat{S}_{k,y}$. (For the other toy example from \cref{sec:qke-vs-multiparticle}, namely a system with two neutrino modes rather than one neutrino mode and one antineutrino mode, only the sign in front of $\hat{S}_{2,z}$ changes. In the following, we will  work with the bipolar neutrino--antineutrino system for definiteness, which is computationally the more challenging of the two examples.) In the $\ket{k_1, \kappa}$ basis, the diagonal entries of this Hamiltonian are
\begin{align}
    \epsilon_{k_1 \kappa} \equiv \bra{k_1,\kappa} \hat{H} \ket{k_1,\kappa}
        = -\Delta\,(m_1 - m_2) + 4 J_{12}\, m_1 m_2 ,
    \label{eq:H-kappa-diag}
\end{align}
where again $m_i = k_i - j$. Similarly, the off-diagonal elements are
\begin{align}
    \bra{k_1+1,\kappa} \hat{H} \ket{k_1,\kappa}
        = 2 J_{12}\, c_{k_1}\, c_{k_2 - 1},
    \label{eq:H-kappa-offdiag}  \\
\intertext{with}
    c_k \equiv \sqrt{(N-k)(k+1)} .
    \label{eq:ck}
\end{align}
Here $c_k$ is the ladder-operator matrix element $\bra{j,m+1}\hat{S}^+\ket{j,m} = \sqrt{(j-m)(j+m+1)}$ which can be expressed in terms of $k=j+m$.
%Correspondingly $c_{k-1} = \bra{j,m-1}\hat{S}^-\ket{j,m}$.
Since $c_N = c_{-1} = 0$, the off-diagonal element vanishes for $k_1 = N$ or $k_2 = 0$, which prevents transitions into unphysical states.

\paragraph{Time Evolution.} To evolve the system under the Hamiltonian defined by \cref{eq:H-kappa-diag,eq:H-kappa-offdiag}, we follow an idea from ref.~\cite{Stryker:2021asy}, namely to split the Hamiltonian into operations that couple pairs of states which differ only in the least significant qubit of register~1, $(\ket{2h,\kappa}, \ket{2h+1,\kappa})$ (``even bonds''), and those which affect pairs of the form $(\ket{2h+1,\kappa}, \ket{2h+2,\kappa})$ (``odd bonds''). This has the advantage that, for even bonds, transitions between the two states $\ket{2h,\kappa}$ and $\ket{2h+1,\kappa})$ affect only a single qubit and are therefore easy to implement. Consequently, we split the Hamiltonian into $\hat{H} = \hat{H}_M + \hat{H}_O$, where $\hat{H}_M$ contains the operations on even bonds as well as the diagonal components, and $\hat{H}_O$ contains the operations on odd bonds.

In each Trotter step, we first apply $\hat{H}_M$. In the subspace spanned by the states $\ket{2h,\kappa}$ and $\ket{2h+1,\kappa}$, that operator is a $2 \times 2$ matrix $m_{h\kappa}$. In terms of the diagonal elements $\epsilon_{k_1\kappa}$ from \cref{eq:H-kappa-diag}, we define their mean and half-difference,
\begin{align}
    \bar\epsilon_{h\kappa}   &\equiv \tfrac12 \big( \epsilon_{2h,\kappa} + \epsilon_{2h+1,\kappa} \big), \quad&
    \delta\epsilon_{h\kappa} &\equiv \tfrac12 \big( \epsilon_{2h,\kappa} - \epsilon_{2h+1,\kappa} \big) .
    \label{eq:eps-bar-delta}
\end{align}
We also use the shorthand notation
\begin{align}
    g_{h\kappa} \equiv \bra{2h+1,\kappa} \hat{H}_M \ket{2h,\kappa} = 2 J_{12}\, c_{k_1}\, c_{k_2 - 1},
    \label{eq:g-even}
\end{align}
for the off-diagonal elements of $\hat{H}_{M}$ from \cref{eq:H-kappa-offdiag}. Note that $k_1 = 2h$ and $k_2 = (\kappa - 2h) \bmod D$.
Defining finally $\Omega_{h\kappa} \equiv \sqrt{\delta\epsilon_{h\kappa}^2 + g_{h\kappa}^2}$, the time evolution over a time interval $\tau$ in the $(\ket{2h,\kappa}, \ket{2h+1,\kappa})$ subspace is described by the operator
\begin{align}
    u_{h\kappa}(\tau) &\equiv \exp(-i\tau\, m_{h\kappa})\\
        &= e^{-i\tau\bar\epsilon_{h\kappa}} \left[
            \cos(\Omega_{h\kappa}\tau) - i\sin(\Omega_{h\kappa}\tau) \,
                \frac{\delta\epsilon_{h\kappa}\, \hat{Z} + g_{h\kappa} \hat{X}}
                {\Omega_{h\kappa}} \right],
  \label{eq:u-h-kappa}
\end{align}
where $\hat{Z}$ and $\hat{X}$ are Pauli matrices acting in the subspace. The full time evolution under $\hat{H}_M$ is then
\begin{align}
    \mathcal{M}(\tau) &\equiv \exp(-i\tau \hat{H}_M) \\[6pt]
        &= \sum_{h,\kappa} \Big( \ket{2h,\kappa}\bra{2h,\kappa}
                      + \ket{2h+1,\kappa}\bra{2h+1,\kappa} \Big)\, u_{h\kappa}(\tau).
  \label{eq:even-bond-circuit}
\end{align}
The sum over $\kappa$ accounts for the (decoupled) $\kappa$ sectors into which the initial state decomposes via~\cref{eq:spin-coherent-expansion}.

As the two states $\ket{2h,\kappa}$ and $\ket{2h+1,\kappa}$ differ only in their least significant qubit, we can implement $\mathcal{M}(\tau)$ as a uniformly controlled $U(2)$ gate following ref.~\cite{Bergholm:2004arh}. A uniformly controlled gate implements a $U(2)$ rotation on the least significant qubit, with the parameters of that rotation depending on all other qubits in the register \cite{Bergholm:2004arh}.  Such an operation is sometimes also referred to as a multiplexer.

We then apply time evolution under the odd half $\hat{H}_O$ of the Hamiltonian by first carrying out a decrement on register~1 (subtraction of $1 \bmod D$), which maps each odd bond $(\ket{2h+1,\kappa}, \ket{2h+2,\kappa})$ onto an even bond $(\ket{2h,\kappa}, \ket{2h+1,\kappa})$. We can then  use the operator $\mathcal{M}(\tau)$ from \cref{eq:even-bond-circuit} again with
\begin{align}
    g_{h\kappa} \to \bra{2h+2, \kappa} \hat{H}_O \ket{2h+1,\kappa}
        = 2 J_{12}\, c_{k_1}\, c_{k_2 - 1},
    \label{eq:g-odd}
\end{align}
where now $k_1 = 2h + 1$ and $k_2 = \kappa - 2h - 1$. Finally, a cyclic increment is applied to register~1 to undo the relabelling.
We implement the cyclic increment as an ancilla-free cascade of multi-controlled $X$ gates \cite{Douglas:2009nqc}. It proceeds from the most significant digit down while flipping each qubit~$p$ conditioned on all less significant qubits in the $\ket{1}$ state, followed by a final $X$ gate on the least significant qubit. The decrement is the inverse gate.

\paragraph{Deferred diagonal block.} The implementation of a uniformly controlled $U(2)$ gate from Ref.~\cite{Bergholm:2004arh} decomposes $\mathcal{M}$ as $\mathcal{M} = \Lambda \mathcal{M}'$, where $\Lambda$ is diagonal in the $D$-dimensional Hilbert space and $\mathcal{M}'$ has a fixed structure of only CNOT gates and single qubit gates. The implementation of $\Lambda$ typically involves both single-qubit and two-qubit gates. It is straightforward to show that $\Lambda.CNOT = CNOT.\tilde\Lambda$, where $\tilde\Lambda$ is also diagonal, though in general $\tilde\Lambda \neq \Lambda$. Similarly, one can show that $\Lambda$ can be commuted past the cyclic increment / decrement operations, past the $W$ transform from \cref{eq:W-transform}, and past any single-qubit gate (at the expense of changing that single-qubit gate). Therefore, all diagonal operators appearing in the circuit can be commuted past the other operators to collect them in a single place, as a single diagonal operator. This leads to an important reduction in circuit depth.

\paragraph{Deployment on quantum hardware.}
We have deployed the algorithm described above on the IBM Kingston QPU, which uses the Heron~R2 architecture. We have chosen the parameters $J_{12} = 2.5 |\Delta| / (2N)$ for the self-coupling, $\theta = 0.05$ for the vacuum mixing angle, and $\tau = 0.25 / |\Delta|$ for the Trotter step size. $|\Delta|$ only sets the overall energy scale and, as long as all other dimensionful parameters are chosen relative to $|\Delta|$, does not enter the circuit. For producing the data points in the top panel of \cref{fig:bipolar}, we have run the circuit with fixed $|\Delta| \tau = 30$ for different total times, sampling the result on the time grid $|\Delta| t = \{30, 60, 90, \ldots, 300\}$. For the middle panel of \cref{fig:bipolar}, we have used $|\Delta| \tau = 0.25$ and we have sampled the result at $|\Delta| t = \{0.25, 0.50, \ldots, 3.5\}$ for the first 14 steps, and $\{4, 4.5, \ldots 6\}$ with twice the spacing for later times. Each run requires a patch of four adjacent qubits for placing the two registers with 3 neutrinos each. To pick the optimal patch based on real-time calibration data, we consider all qubits with readout error below 5\% and all connections with CZ error below 1\%. Among this set, we choose the patch with the lowest estimated error rate for the deepest circuit under consideration. For error mitigation, we have used the Qiskit Runtime Estimator class at resilience level 2. This includes Pauli twirling for gates and measurements \cite{Wallman:2015uzh, PhysRevA.105.032620}, zero-noise extrapolation at noise factors 1, 1.5, and 2 with a linear fit \cite{Temme:2016vkz, 9259940}, and XY4 dynamic decoupling \cite{PhysRevLett.82.2417}. For each noise factor, single-qubit pre-measurement rotations are applied to all qubits to measure the expectation value along each Pauli basis ($X$, $Y$, and $Z$) with 2\,048 shots each. Each time point therefore uses $3 \times 3 \times 2048$ shots in total. Over $|\Delta|t \leq 3.5$, the root-mean-square error of the mitigated observable deviates from the exact solution by 0.053 in $P_\nu$ and $0.068$ in $P_{\bar\nu}$. (Without zero-noise extrapolation, these errors increase to 0.077 and 0.098.)

To also quantify the Trotterization error, we compare a noiseless implementation of the quantum circuit on classical hardware to the exact solution, obtained by directly exponentiating the full Hamiltonian. We find that after $n_{\rm steps} = 4$ Trotter steps, the largest Trotterization error on the survival probability across the two modes is $7.3\times10^{-3}$, $0.043$, and $0.33$ for $N = 3, 7, 15$, respectively. Reducing the Trotter step size by using $n_{\rm steps} = 64$ and $\tau = 1/(16 |\Delta|)$ results in reduced relative errors of \num{2.1e-5}, \num{7.4e-6}, and \num{7.1e-4}.

We finally emphasize that the main advantage of the algorithm presented here is not necessarily an extremely low number of two-qubit gates. In this regard, it is still a significant improvement compared to the algorithms presented in \cite{Bleau:2026iiq}, and requires a similar number of two-qubit gates as a naive implementation where each neutrino is represented by one qubit. The main strength of our new algorithm is that it requires only a small number of qubits to simulate a large number of neutrinos. It is therefore best suited for QPU architectures with low noise, but only few qubits. In that sense, deployment on IBM's transmon QPUs, which have the opposite characteristic, was not optimal.

\paragraph{Generalizations.}
We briefly mentioned above that the algorithm introduced here is generalizable to systems with more than two modes. To achieve this, it is no longer sufficient to apply the $W$ transform, which reduces the interaction between the two modes to an operation on a single register, only once at the beginning of the circuit. Rather, each interaction between two modes needs to be wrapped between $W$ and $W^\dag$ operations that combine the registers of those two modes. For a system with $n$ modes, this implies that each Trotter step contains $n(n-1)/2$ pairs of $W$ and $W^\dag$ transformations. The full circuit would correspond to the gate sequence in \cref{fig:circuit} being applied between all register pairs in every Trotter step.

%-----------------------------------------------------------------------------
\subsection{Results: Emergence of Quantum Kinetic Behavior in the Large-$N$ Limit}
\label{sec:results-qke-vs-multiparticle}
%-----------------------------------------------------------------------------

We are now ready to return to \cref{fig:bipolar} and compare the time evolution of the full multiparticle system to the predictions of the quantum kinetic (mean field) approach. In the figure, the solid black line corresponds to the mean-field prediction obtained by solving \cref{eq:von-Neumann}. Lines colored in different shades of purple were computed using multiparticle dynamics on classical hardware, along the lines of \cref{sec:bipolar-algo}. The results of the quantum circuits from \cref{sec:quantum-algo}, finally, are plotted in blue, with data points corresponding to runs on the IBM~Kingston quantum processor, and the blue line showing results from the same circuit on a noiseless simulator on classical hardware.

As expected from our analytical results in \cref{sec:qke-emergence}, we find that the mean-field and multiparticle approaches agree in the large-$N$ limit, while showing significant differences for small systems. For the neutrino--neutrino toy system shown in the top panel of \cref{fig:bipolar}, we see that $N \gtrsim 300$ is already sufficient to achieve excellent agreement. For the neutrino--antineutrino system in the middle panel, on the other hand, the same level of convergence requires $N \gtrsim 4\,000$. The main reason for this is the different choice of the self-interaction strength, parameterized here as $2 N J_{12} / \Delta_1$. We can understand this from the analytical results in \cref{sec:qke-emergence}: without vacuum oscillations, the states $\nu_A$ and $\bar\nu_X$ from \cref{eq:nu-A,eq:nu-X-bar} coincide with flavour eigenstates. All the flavour-changing terms in \cref{eq:ev-alpha} that are coherently enhanced (the second term in the first line and all the terms on the second line) then come with small coefficient, suppressed by factors of the form $\sprod{\alpha}{A}$ and/or $\sprod{\bar{Y}}{\bar{A}}$. This makes it easier for the last, non-coherent term to compete. Only after vacuum oscillations have driven $\nu_A$ and $\bar\nu_X$ away from the flavour eigenbasis, coherent enhancement comes into play.

Note that, besides the smaller interaction strength, we have also chosen $\Delta_1 \neq \Delta_2$ for the $\nu_e$--$\nu_\mu$ system in the top panel of \cref{fig:bipolar}. Without this small asymmetry, we would have found essentially vacuum oscillations. This can be understood in the flavour polarization vector formalism (see \cref{sec:bipolar}): the $\nu_e$ and $\nu_\mu$ initial state correspond to polarization vectors of equal length and pointing in opposite directions, so the net polarization would be zero. If the two polarization vectors precess around $\hat{\vec{b}}$ at the same speed (i.e., $\Delta_1 = \Delta_2$), this will never change. Only for $\Delta_1 \neq \Delta_2$, they can become imbalanced, leading to an enhanced flavour transition amplitude.

Comparing results derived on classical hardware to those obtained on quantum hardware, we see that the noisy quantum computation faithfully reproduces the classical result for about 10 Trotter steps before noise become visible. Results remain meaningful for up to 20 Trotter steps, beyond which the simulation becomes entirely noise-dominated. We have checked that when simulating a larger system ($N=7$ instead of $N=3$), the number of meaningful Trotter steps reduces to $\sim 8$.

The take-home message from \cref{fig:bipolar} is that full multiparticle simulations of collectively oscillating neutrino systems reproduce the predictions of the quantum-kinetic mean-field approach in the large-$N$ limit. We have explicitly demonstrated this for simple toy systems, and in the process have taken significant steps forward in developing new algorithms for large multiparticle systems on both classical and quantum hardware. It stands to reason that the conclusions derived by our two-mode toy systems generalize to more complex systems with many modes, \emph{as long as there are all-to-all interactions}. We will investigate in the next section how our conclusions change for systems in which interactions are not all-to-all.

%=============================================================================
\section{Phenomenologically Relevant Entanglement}
\label{sec:systems-with-entanglement}
%=============================================================================

Do the arguments given above show conclusively that quantum entanglement is irrelevant to collective neutrino oscillations in realistic astrophysical and cosmological environments? We now argue that they do not. We recall that the main arguments against phenomenologically relevant quantum entanglement are:
\begin{enumerate}
    \item In a realistic environment, neutrinos are streaming in different directions, therefore each pair of neutrinos sees each other at most once. This prevents the build-up of significant entanglement.

    \item Even for toy systems with persistent all-to-all interactions, we have seen that, in the large $N$ limit, a quantum kinetic description is sufficient since only processes that do not generate entanglement get coherently enhanced.
\end{enumerate}

%-----------------------------------------------------------------------------
\subsection{Neutrino Streams}
\label{sec:neutrino-chain}
%-----------------------------------------------------------------------------

\begin{figure}
    \centering
    \includegraphics[width=\linewidth]{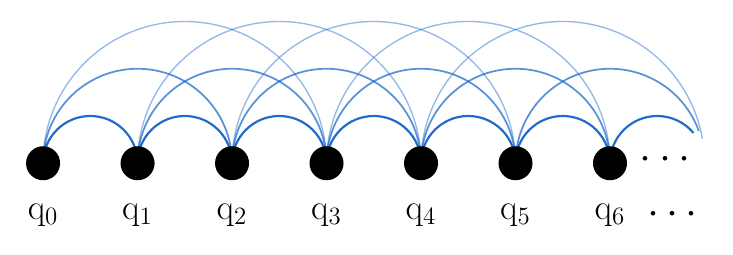}
    \caption{Setup for a quantum simulation of a stream of neutrinos leaving a supernova, with partially overlapping wave packets. Each qubit $q_0$, $q_1$, \ldots\ (black dots) encodes the flavour state of one neutrino. Neutrinos have sizeable interactions with their nearest neighbours and weaker interactions with their next-to-nearest and next-to-next-to-nearest neighbours (blue arcs).}
    \label{fig:neutrino-chain-setup}
\end{figure}

To see entanglement, we are therefore looking for a system in which only a small number of neutrinos get entangled, and in which these neutrinos interact with each other for an extended period of time.
Consider an ensemble of neutrinos traveling in a particular direction. If the neutrino density is extremely high, and the size of each neutrino wave packet is larger than the separation between neutrinos, interactions become possible. In particular, each neutrino will interact with its nearest neighbours (that is, the neutrinos directly ahead of it and directly behind it). Longer-range interactions will be suppressed exponentially for Gaussian wave packets or as a power law for Lorentzian wave packets. The scenario is schematically depicted in \cref{fig:neutrino-chain-setup}. As all neutrinos travel at the same speed, these interactions are active for an extended period of time. They only cease when minor kinematic differences eventually drive the neutrinos apart.

To gauge whether this scenario can be realized in nature, for instance in neutrinos streaming away from a supernova or a neutron star merger, we need to examine the width of a typical supernova neutrino wave packet, which has been estimated to be of order \SI{e-12}{cm} \cite{Akhmedov:2017mcc} (see also ref.~\cite{Kersten:2015kio} for a very similar estimate). This is comparable to the mean distance between neutrinos at typical densities around the supernova neutrinosphere, where $n_\nu \sim \SI{e38}{cm^{-3}}$ implies a mean separation of order \SI{e-13}{cm}. However, neighbouring neutrinos will not typically stream in the same direction. Therefore, according to these estimates, the time during which the interactions depicted in \cref{fig:neutrino-chain-setup} are active will be extremely short, leaving insufficient time to build up entanglement.

This conclusion can be avoided if neutrino wave packets are significantly larger than estimated in Refs.~\cite{Kersten:2015kio, Akhmedov:2017mcc}. We speculate here that this could indeed be the case thanks to the presence of strong magnetic fields in supernova cores and neutron star mergers. In a magnetic field, charged particles experience Landau quantization, so it stands to reason that nucleons and charged leptons are delocalized across their Landau levels. The spatial extent of a Landau level is comparable to the cyclotron radius, which for a particle of charge $e$ and momentum $p$ in a magnetic field of strength $B$ is given by $r_c = p / (e \, B) \simeq \SI{1}{nm} \times (p / \SI{100}{MeV/c}) \times (B / \SI{e12}{G})^{-1}$. Neutrinos produced from such a system will inherit this delocalization. At the aforementioned densities of order \SI{e38}{cm^{-3}}, each neutrino wave packet of size \SI{e-7}{cm} will then overlap with $10^{17}$ other neutrinos. The likelihood that some of these neutrinos have momentum vectors sufficiently close to the one of the test neutrino to retain wave function overlap over macroscopic distances is now of order one. There is then sufficient time to build up phenomenologically relevant quantum entanglement among neutrinos, and the one-dimensional neutrino stream from \cref{fig:neutrino-chain-setup} can serve as a good proxy to understand the physical consequences.

These considerations motivate us to push ahead with the simulation of one-dimensional neutrino streams and to set up a quantum simulation of the system shown in \cref{fig:neutrino-chain-setup}. Each black dot in the diagram now represents a qubit encoding the flavour of one neutrino. The interactions between qubits have the same structure as in \cref{eq:Hqnunu}, but the coupling constants $J_{pq}$ are chosen such that the $j$-th qubit has the strongest interactions with qubits $j-1$, $j+1$ (representing the neutrinos just ahead of it and just behind it), and progressively weaker interactions with more distant qubits. Specifically, we choose the interaction strength between qubits $p$ and $q$ as
\begin{align}
    g_{pq} = \begin{cases}
                 N_{\mathrm{qubits}} / |p-q|  & \text{for $|p-q| \leq d_{\rm max}$} \\
                 0                            & \text{otherwise}
             \end{cases} ,
    \label{eq:g-pq}
\end{align}
where $d_{\rm max}$ imposes a cutoff on interactions between qubits that are too far apart. This cutoff is introduced to reduce the number of two-qubit gates required in a quantum simulation.

\begin{figure}
    \centering
    \includegraphics[width=\linewidth]{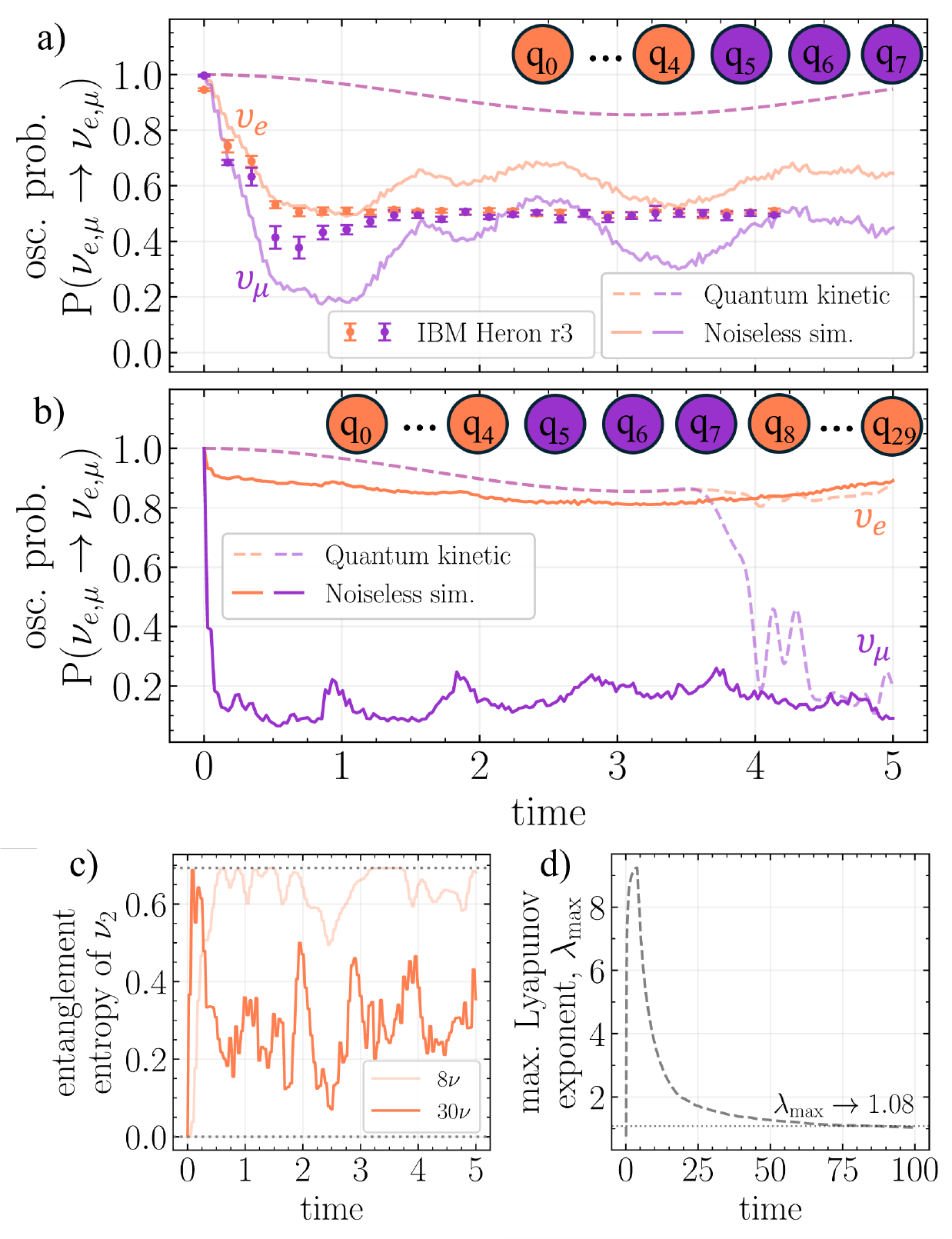}
    \caption{
    Average survival probability of electron (orange) and muon (purple) neutrinos in a one-dimensional neutrino stream (cf.\ \cref{fig:neutrino-chain-setup}) as a function of time. In panel (a) we consider an 8-neutrino stream and compare the solution from the quantum kinetic mean-field approach, where in the self-interaction term in \cref{eq:von-Neumann}, the integral is replaced by a weighted sum over just the nearest and next-to-nearest neighbours (dashed curves), a noiseless simulation on classical hardware (solid curves), and a noisy simulation on an IBM Heron r3 QPU, specifically the IBM Aachen device (data points with error bars).
    The three rightmost neutrinos in the chain are initially $\nu_\mu$, the rest are initialized as $\nu_e$. Panel (b) corresponds to a 30-neutrino system with interactions up to next-to-next-to-nearest neighbours, for which we only show the mean-field and noiseless simulations as simulations on real hardware become noise-dominated too quickly.  Again, three neutrinos are initially taken in the $\ket{\nu_\mu}$ state, the rest in the $\ket{\nu_e}$ state, as shown in the diagram inset into the plot.
    We see that, even with a larger number of neutrinos, the quantum simulation retains features which are lost in the quantum kinetic approach.
    Panel (c) shows the entanglement entropy as a function of time for the third neutrino in the stream in the noiseless simulation, proving the build-up of significant entanglement.
    Panel (d) shows the maximal Lyaponov exponent as a function of time for the quantum kinetic solution, where a positive value points towards the emergence of chaotic behaviour, which we indeed observe in panel (b).
    }
    \label{fig:neutrino-chain-results}
\end{figure}

In \cref{fig:neutrino-chain-results}, we compare the average survival probability of electron and muon neutrinos in a neutrino chain over time, calculated using the quantum kinetic mean-field approach, a noiseless quantum simulation on classical hardware, and a noisy quantum simulation on an IBM Heron~r3 QPU, specifically the IBM Aachen device. For the quantum simulations, the system's time evolution is described by the Hamiltonian in \cref{eq:Hq0,eq:Hqnunu}, where we set $\Delta_p = 1$ for all neutrinos, $\theta = 0.195$ and $J_{pq} = \sqrt{2}G_F g_{pq}$, with $G_F = 0.25$.
In panel (a) of \cref{fig:neutrino-chain-results}, we have chosen $d_{\rm max} = 2$ (interactions up to the next-to-nearest neighbour) to keep noise manageable; in panel (b), we have used $d_{\rm max} = 3$ (interactions up to the next-to-next-to-nearest neighbour). We choose an initial state in which the sixth, seventh and eighth neutrino along the chain are muon neutrinos and the rest are electron neutrinos. In the quantum kinetic approach (dashed curves in \cref{fig:neutrino-chain-results}), we solve \cref{eq:von-Neumann}, neglecting the spatial transport term $\vec\nabla_{\vec{x}} \rho_{\vec{x},\vec{p},t}$, and replacing the integral $\int \! d^3p' \, (1 - \vec{\hat{p}} \cdot \vec{\hat{p}}') \rho_{\vec{p'},t}$ by a sum $\sum_{|p-q| \leq d_{\rm max}} g_{pq} \rho_{\vec{p},t}$, with the coupling constants $g_{pq}$ given by \cref{eq:g-pq}. For the quantum simulations, we use the approach described in \cref{sec:implementation}, which consists of evolving the Hamiltonian in \cref{eq:Hq0,eq:Hqnunu} in time using the second-order Suzuki--Trotter decomposition in \cref{eq:Trotter}. In the noiseless simulation, we use shot-based sampling with 1\,024 shots and the \texttt{AerSimulator()} class implemented in Qiskit \cite{javadiabhari2024quantumcomputingqiskit} with the matrix product state (MPS) method.\footnote{For an 8-qubit chain, which is small enough to also be simulated using the state vector and density matrix methods, we have checked explicitly that the three methods agree very well. This demonstrates that the assumption of negligible long-range entanglement, which is at the core of the MPS method, is valid in our case.} For the noisy simulation on superconducting quantum hardware, we computed our results using the \texttt{Estimator} primitive and used a Heron~r3 IBM QPU with error mitigation through Pauli twirling \cite{Wallman:2015uzh}, dynamical decoupling \cite{PhysRevLett.82.2417} with an XpXm pulse sequence, zero-noise extrapolation \cite{Temme:2016vkz, 9259940} and twirled readout error extinction \cite{PhysRevA.105.032620}.

The accuracy of the simulation on the QPU is limited for this system due to the large number of noisy two-qubit operations necessary to account for interactions between neutrinos up to the next-to-nearest neighbour. We can decrease the number of two-qubit gates by limiting the number of qubits and by increasing the size of the Trotter steps, albeit the latter increases the Trotterization error. In panel (a) of \cref{fig:neutrino-chain-results}, we have balanced the number of two-qubit gates against the size of the Trotter steps such that the noisy simulation resolves the first oscillation period. At later times, though, the simulation reaches $\sim 200$ two-qubit gates, causing the system to become noise-dominated and all oscillation probabilities to converge towards 0.5.

As the number of neutrinos increases, such as in panel (b) of \cref{fig:neutrino-chain-results} for a 30-neutrino system, we find that the quantum simulation (which in this case is only feasible using the MPS simulator on classical hardware) still retains features which are lost in the quantum kinetic approach. This shows that the quantum kinetic approach is indeed not able to model collective neutrino oscillations in a neutrino stream, providing evidence for the arguments motivating quantum simulations above. To further corroborate this conclusion, we compute the entanglement entropy of the system over time. The entanglement entropy for a single neutrino is given by
\begin{align}
    S(t) \equiv -\sum_i \lambda_i(t) \log\lambda_i(t) ,
\end{align}
where $\lambda_i$ are the eigenvalues of the reduced density matrix of the neutrino under consideration, and where the degrees of freedom of all other neutrinos have been traced out. The result, shown in \cref{fig:neutrino-chain-results}~(c) for the third neutrino in the chain, demonstrates that entanglement does indeed build up. We have checked that the entanglement entropy behaves similarly for the other neutrinos except those at the edges.

Another notable point is the possible emergence of chaotic behaviour in the solution of the quantum kinetic equations. Indeed, we see in the right part of \cref{fig:neutrino-chain-results}~(b) that, after $t \simeq 2.5$, the solution becomes irregular. We have checked that the solution does not stabilize as we vary the tolerances and step sizes of the ODE solver. To see whether the behaviour is indeed chaotic, we have calculated the maximal Lyapunov exponent, $\lambda_{\mathrm{max}}$, using a two-trajectory Benettin algorithm and found that $\lambda_{\mathrm{max}}$ converges to a positive value, which points towards a chaotic solution \cite{Hansen:2014paa}, see \cref{fig:neutrino-chain-results}~(d). The algorithm consists of constructing a physical perturbation, in this case a small rotation of the density matrix of one of the neutrinos in the stream, and measuring the separation $d = \sqrt{\tr[(\rho_{\rm ref} - \rho_{\rm pert})^\dagger(\rho_{\rm ref} - \rho_{\rm pert})]}$ between the perturbed and unperturbed (reference) solutions. The reference and perturbed trajectories are evolved independently and the separation $d$ is accumulated over time steps until a chosen end time, $t_{\rm final}$, by which the Lyaponov exponent should have converged towards a final value. The maximal Lyaponov exponent is thus given by 
\begin{align}
    \lambda_{\rm max} = \frac{1}{t_{\rm final}} \log\left(\frac{d}{d_{\rm initial}}\right),
\end{align}
where a positive $\lambda_{\rm max}$ means that initially nearby trajectories diverge exponentially, the hallmark of a chaotic solution.

%-----------------------------------------------------------------------------
\subsection{``Once-in-a-Lifetime Encounter" Argument}
\label{sec:once-in-a-lifetime}
%-----------------------------------------------------------------------------

We have argued that phenomenologically relevant entanglement may be realized in the special case of a neutrino stream being radiated in a particular direction. However, in the case where neutrinos are streaming in different directions, they will not interact with each other more than once, and therefore may not become strongly entangled. We now show the limitations of this ``once-in-a-lifetime encounter'' argument.

\begin{figure}
    \centering
    \includegraphics[width=\linewidth]{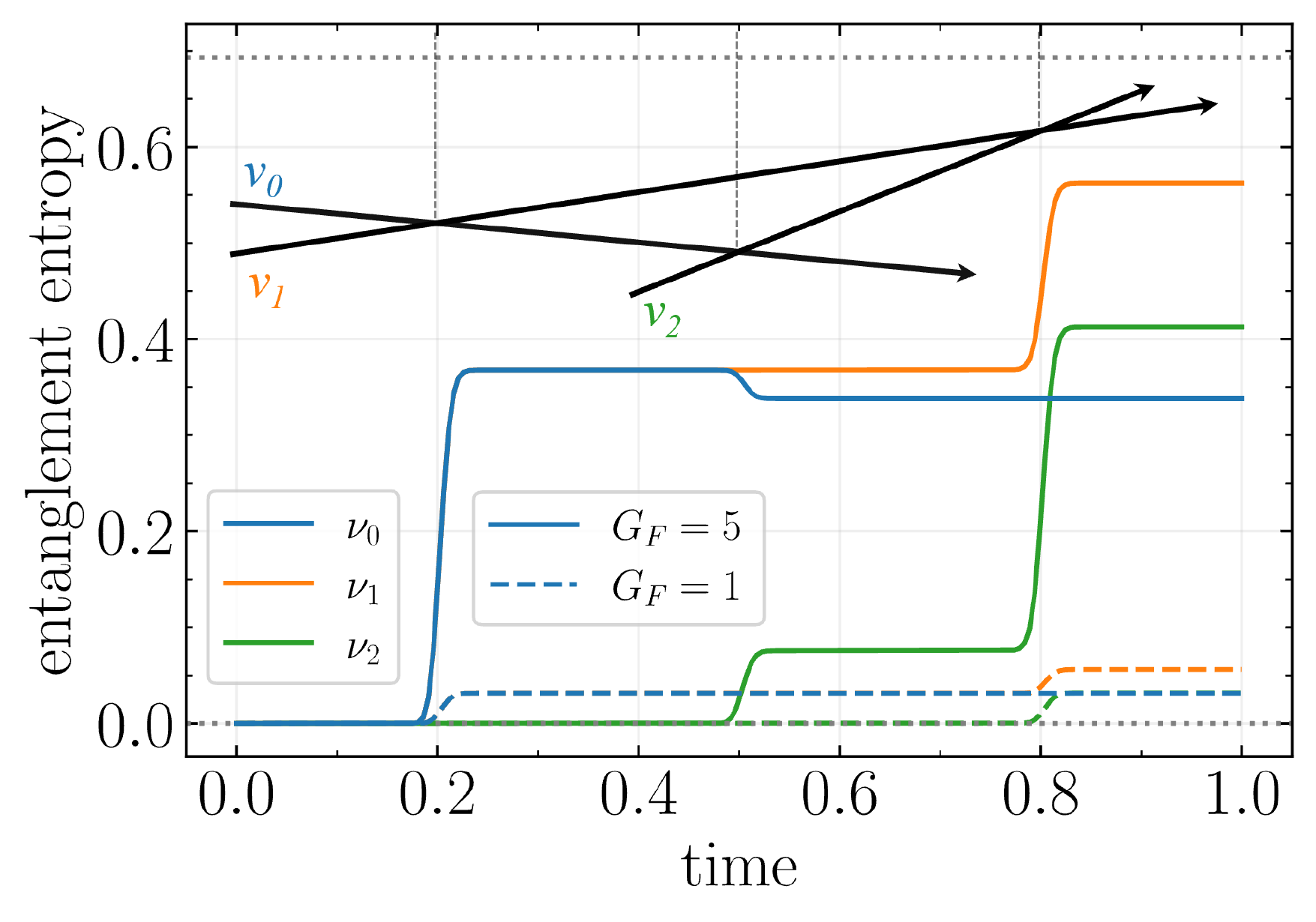}
    \caption{Entanglement entropy as a function of time for a toy system of three neutrinos, where each pair of neutrinos only interacts once, at times $t=(0.2,\;0.5,\;0.8)$. The entanglement build-up increases with the coupling strength $G_F$. Consequently, in the realistic case of small couplings, entanglement effects are negligible.}
    \label{fig:3-nu-results}
\end{figure}

We consider first a toy system of three neutrinos ($\nu_0$, $\nu_1$, $\nu_2$) as depicted in the inset of \cref{fig:3-nu-results}, where $\nu_1$ is initially a muon neutrino and the rest are electron neutrinos, and each pair interacts only once. The system's density matrix is evolved in time by solving its von Neumann equation with the Hamiltonian (cf.\ \cref{eq:H-q,eq:Hq0,eq:Hqnunu})
\begin{align}
    H = \sum_{p=0}^2 H_0^{(p)}
      + \sqrt{2} G_F \sum_{p<q}
        \exp\bigg[\!-\frac{(t-t_c^{pq})^2}{2\tau_{\rm int}^2} \bigg] \,
        \hat{\vec\sigma}_p \cdot \hat{\vec\sigma}_q,
    \label{eq:H-oial}
\end{align}
where
\begin{align}
    H_0^{(p)} = \Delta \big( \sin(2\theta) \hat\sigma^x_p
                           - \cos(2\theta) \hat\sigma^z_p \big).
\end{align}
The Gaussian in the interaction term in \cref{eq:H-oial} accounts for the limited duration of each encounter by restricting each interaction to a time window of width $\tau_{\rm int}$ centered at a time $t = t_c^{pq}$. For the numerical evaluation, we choose the parameters $\Delta = 1$, $\theta = 0.195$, and $\tau_{\rm int} = 0.01$. After tracing out $\nu_1$ and $\nu_2$, we calculate the entanglement entropy of $\nu_0$, $S(\rho_0) = -\sum_i \lambda_i^0 \textrm{log} \lambda_i^0$, where $\lambda_i^0$ are the eigenvalues of the reduced density matrix $\rho_0$ of $\nu_0$. We calculate the entanglement entropies of $\nu_1$ and $\nu_2$ analogously.

Our findings are summarized in \cref{fig:3-nu-results}. For an initial state consisting of two electron neutrinos and one muon neutrino, we find that the entanglement entropy of all three neutrinos increases over time. This means that a build-up of entanglement \textit{is} possible. However, as we decrease the coupling strength $G_F$, the build-up becomes much less prominent. Therefore, for realistic couplings between neutrinos in a supernova environment, we do not expect multiple once-in-a-lifetime encounters to lead to phenomenologically relevant degrees of entanglement. We also note that interactions are sparser in a three-dimensional system than in our planar toy system, since in three dimensions, most pairs of neutrino trajectories never cross.

\begin{figure}
    \centering
    \includegraphics[width=\linewidth]{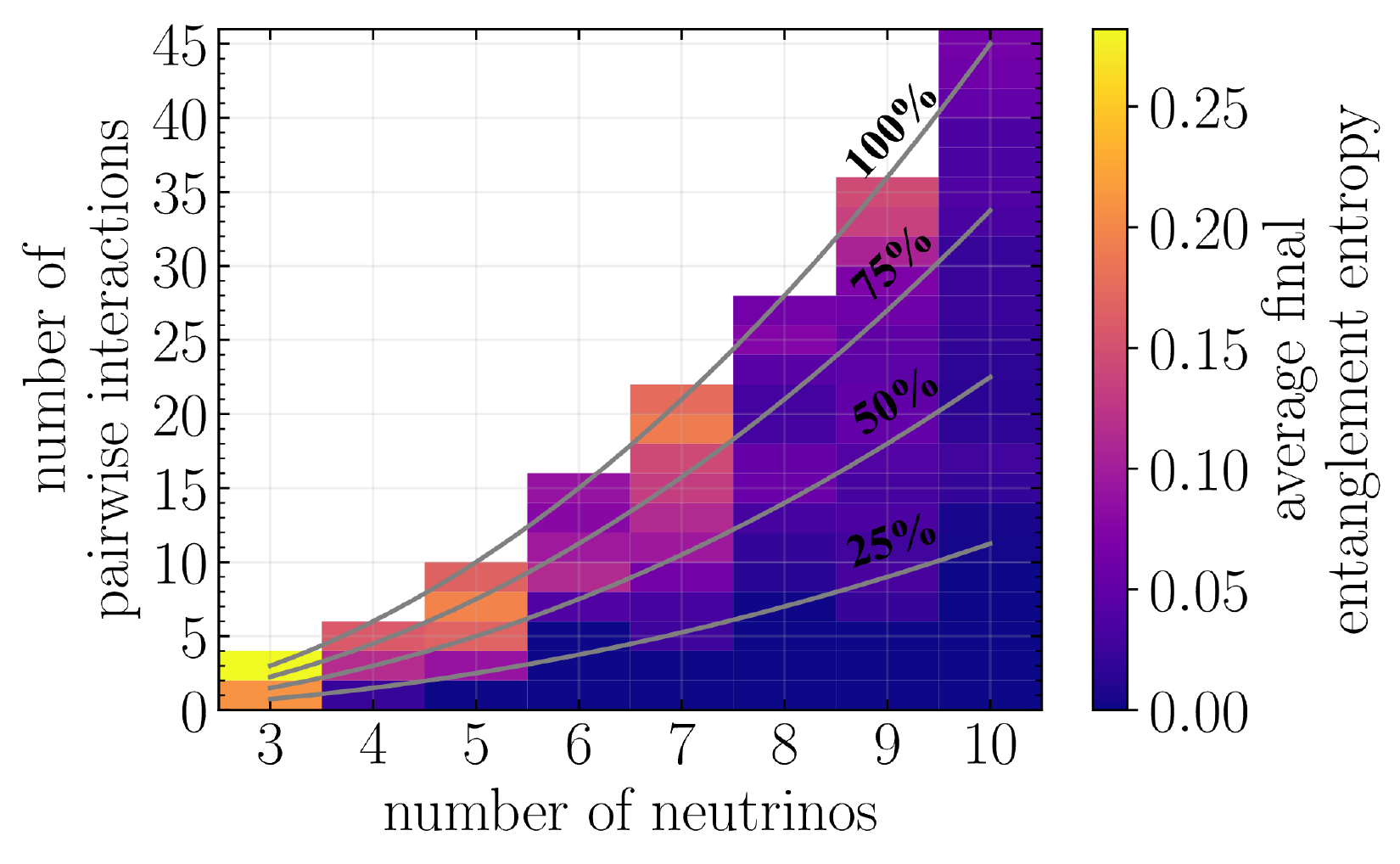}
    \caption{Average entanglement entropy over all neutrinos in an system with once-in-a-lifetime interactions as a function of the system size (number of neutrinos) and the number of pairwise interactions. We set $G_F = 5$ and choose an initial state consisting of half electron and half muon neutrinos. We see that few neutrinos or many pairwise interactions lead to a larger entanglement entropy. The grey curves represent different fractions of the maximum number of pairwise interactions.}
    \label{fig:N-nu-results}
\end{figure}

As a generalization of this discussion, we increase the size of the system to up to ten neutrinos, leaving the number of pairwise interactions as a free parameter, and choosing the participating neutrino pairs at random with the timing of the interactions evenly spaced between $t=0$ and $t=1$. We forbid repeated interactions of the same neutrino pair. In \cref{fig:N-nu-results}, we show the average entanglement entropy over all neutrinos in the system after a given number of pairwise interactions has been completed. Unsurprisingly, a larger number of interactions leads to a larger entanglement entropy. Large entanglement entropies are easier to reach in small systems than in large ones because, for a fixed total number of interactions, each neutrino experiences more interactions in a small system than in a large one. Extrapolating these trends towards supernova conditions, i.e., a system with a large number of neutrinos that allows for a large number of pairwise interactions, we find that the build-up of entanglement from the many total interactions would not counterbalance the small entanglement entropy of each individual neutrino. In other words, we expect the larger number of neutrinos in a supernova to strongly hinder the build-up of entanglement.

%=============================================================================
\section{Conclusions}
\label{sec:conclusions}
%=============================================================================

To recapitulate, we have investigated the possible phenomenological impact of quantum entanglement among the flavours of different neutrinos in a dense neutrino gas, with supernovae, neutron star mergers, and the early Universe in mind. Our results can be summarized as follows.
\begin{enumerate}
    \item \textbf{In $N$-neutrino toy systems with all-to-all interactions, the dynamics converge to the mean-field limit at large $N$.} We have demonstrated this explicitly for a system consisting initially of similar numbers of $\nu_e$ and $\nu_\mu$, as well as for a system consisting of equal numbers of $\nu_e$ and $\bar\nu_e$. To corroborate our analytic results, we have carried out multiparticle simulations with up to \num{2.6e6} neutrinos on classical hardware and up to 6 neutrinos on a superconducting quantum processor. Our results are summarized in \cref{fig:bipolar}. In both cases, we have developed highly optimized algorithms that allow us to go to large $N$ and, in the case of the simulation on noisy hardware, simulate more than 10 Trotter steps before becoming noise-dominated.

    \item \textbf{Quantum entanglement can become phenomenologically relevant in systems where interactions have a short range (i.e.\ no all-to-all connectivity), but still a long duration.} As an example, we have analyzed a linear stream of neutrinos using again both classical and quantum hardware. Our results, presented in \cref{fig:neutrino-chain-results}, clearly reveal discrepancies between the full multiparticle treatment and the computationally cheaper mean-field approximation. We have contemplated the possibility that such a system may be realized in environments with large magnetic fields if the neutrinos' parent particles are delocalized over Landau levels.

    \item \textbf{If neutrinos only experience short-duration interactions, with each pair meeting at most once (``once-in-a-lifetime encounters''), entanglement is unlikely to build up.}
\end{enumerate}
Our results show that answering the question of whether multiparticle dynamics is phenomenologically relevant in astrophysical systems, and therefore whether quantum computing may be needed to understand these systems, requires a more detailed understanding of the relevant neutrino production processes and, in particular, of the wave packet properties of the resulting neutrino states.

%=============================================================================
\section*{Acknowledgments}
%=============================================================================

It is a pleasure to thank Ting Cheng, Basudeb Dasgupta, Enrique Rico Ortega, and Cristina Volpe for insightful and illuminating discussions. We are moreover grateful to CERN, where crucial parts of this work were carried out, for hospitality and for access to IBM QPUs through the \href{quantum.cern}{CERN Quantum Technology Initiative}. We moreover acknowledge use of the Fir cluster at Simon Fraser University as part of the Digital Research Alliance of Canada. KB and JK have received support from the Cluster of Excellence \href{https://www.prisma.uni-mainz.de/}{``Precision Physics, Fundamental Interactions and Structure of Matter``}, PRISMA${}^{++}$ (EXC 2118/2, Project ID 390831469) funded by the German Research Foundation (DFG). UR would like to acknowledge support from the Department of Atomic Energy, Government of India (Project Identification Number RTI 4002) and from the J.~C.\ Bose Grant of the Anusandhan National Research Foundation (ANRF), Government of India (ANRF/JBG/2025/000265/PS).

%=============================================================================
\appendix
\section{Hamiltonian Matrix Elements of the Bipolar System}
\label{sec:H-bipolar-explicit}
%=============================================================================

In this appendix, we derive the matrix elements of the Hamiltonian describing a bipolar system, \cref{eq:H-bipolar-diag-final,eq:H-bipolar-offdiag-final}. Our starting point is \cref{eq:Hs-n-modes}, in which we drop the self-interaction term as before, so that the Hamiltonian becomes
\begin{align}
    \hat{H} \equiv \vec{b}_1 \cdot \hat{\vec{S}}_1 + \vec{b}_2 \cdot \hat{\vec{S}}_2
             + 4 J_{12} \hat{\vec{S}}_1 \cdot \hat{\vec{S}}_2 .
    \label{eq:Hs-2-modes-appendix}
\end{align}
In a slight generalization of the discussion in \cref{sec:bipolar-algo}, we allow $\vec{b}_1$ and $\vec{b}_2$ to differ in length and orientation and write $\vec{b}_{1,2} \equiv \Delta_{1,2} \hat{\vec{b}}$. The following results are therefore applicable to both systems shown in \cref{fig:bipolar}: the neutrino--neutrino system as well as the neutrino--antineutrino system. After transforming to the primed basis from \cref{sec:bipolar-algo} in which the $z$-axis is aligned with $\hat{\vec{b}}$ and defining
\begin{align}
    \vec{D} &\equiv \vec S_1' - \vec S_2' ,
    \qquad
    \bar \Delta \equiv \frac{\Delta_1 + \Delta_2}{2},
    \qquad
    \delta \Delta \equiv \frac{\Delta_1 - \Delta_2}{2}.
\end{align}
\cref{eq:Hs-2-modes-appendix} becomes
\begin{align}
    \hat{H} = \hat{H}_{\rm vac} + \hat{H}_{\rm int}
    \label{eq:bipolar-H}
\end{align}
with
\begin{align}
    \hat{H}_{\rm vac} \equiv \bar\Delta \, \hat{S}_{{\rm tot}, z}
                           + \delta\Delta \, \hat{D}_z ,
    \qquad
    \hat{H}_{\rm int} \equiv 4 J_{12} \hat{\vec{S}}_1 \cdot \hat{\vec{S}}_2 
    \label{eq:bipolar-Hvac-Hint}
\end{align}
Note that for the special case $\Delta_1 = -\Delta_2$ relevant for bipolar oscillations, $\bar\Delta$ vanishes.

We will work in a basis of $\hat{\vec{S}}_{\rm tot}^2$ and $\hat{M}'$ eigenstates, $\ket{J s_1 s_2 M}$, where $J (J+1)$, $s_1(s_1+1)$, $s_2(s_2+1)$, and $M$ are eigenvalues of \smash{$\hat{\vec{S}}_{\rm tot}^2$}, \smash{$\hat{\vec{S}}_1^2$}, \smash{$\hat{\vec{S}}_2^2$}, and $\hat{M}'$, respectively. Here and in the remainder of this appendix, we denote the eigenvalues under $\hat{M}'$ simply as $M$ (without a prime) to make the notation less cumbersome. It is, however, understood that the quantization axis is always the primed one. We will moreover use the shorthand notation $\ket{J M} \equiv \ket{J s_1 s_2 M}$. We will see that \smash{$\hat{H}_{\rm int}$} is diagonal in this basis, while \smash{$\hat{H}_{\rm vac}$} leaves only $M$ invariant but can change $J$. We discuss $\hat{H}_{\rm int}$ first.

%-----------------------------------------------------------------------------
\subsection{Interaction Term}
%-----------------------------------------------------------------------------

We can rewrite $\hat{H}_{\rm int}$ from \cref{eq:bipolar-Hvac-Hint} as
\begin{align}
    \hat{H}_{\rm int} = 2 J_{12} \big( \hat{\vec{S}}_{\rm tot}^2
        - \hat{\vec{S}}_1^2 - \hat{\vec{S}}_2^2 \big) ,
    \label{eq:H-bp-int}
\end{align}
which is evidently diagonal in the $\ket{J s_1 s_2 M'}$ basis. The diagonal matrix elements are
\begin{multline}
    4 J_{12} \amp{JM}{\hat{\vec{S}}_1 \cdot \hat{\vec{S}}_2}{JM} \\
        = 2 J_{12} \big[ J(J+1) - s_1(s_1+1)-s_2(s_2+1) \big].
    \label{eq:me-bipolar-int}
\end{multline}

%-----------------------------------------------------------------------------
\subsection{Vacuum Term -- Diagonal Part}
%-----------------------------------------------------------------------------

Looking at \cref{eq:bipolar-Hvac-Hint}, it is evident that the first term in $\hat{H}_{\rm vac}$, namely $\bar\Delta \, \hat{S}_{{\rm tot},z}$, is diagonal in the $\ket{J s_1 s_2 M'}$ basis, with matrix elements
\begin{align}
    \bra{JM} \bar\Delta \, \hat{S}_z \ket{JM}
      = \bar\Delta M .
    \label{eq:me-bipolar-vac-S}
\end{align}
The second term, $\delta\Delta \, \hat{D}_z$, on the other hand, has both diagonal and off-diagonal entries. To evaluate the matrix elements of this term, we first apply the Wigner--Eckart theorem, which states that \cite{Varshalovich:1988ifq, Sakurai:2011zz}
\begin{multline}
    \bra{J'M'} \hat{D}_z \ket{JM}
    =
    (-1)^{J'-M'}
    \begin{pmatrix}
        J' & 1 & J \\
        -M' & 0 & M
    \end{pmatrix}
    \langle J' \| \hat{\vec{D}} \| J \rangle ,
    \label{eq:d-prime-wigner-eckart}
\end{multline}
thereby defining the reduced matrix elements $\langle J' \| \hat{\vec{D}} \| J \rangle$. The fact that the Wigner--Eckart theorem can be applied to the components of $\hat{\vec{D}}$ follows from the fact that $\hat{\vec{D}}$ transforms as a vector operator, which can be formally shown by demonstrating that $[\hat{S}_{\rm{tot},a}, \hat{D}_b] = i \epsilon_{abc} \hat{D}_c$~\cite{Sakurai:2011zz}. We see from \cref{eq:d-prime-wigner-eckart} that the matrix elements are non-zero only when $M'=M$ and $J' = J, J\pm1$. This shows that $\hat{H}_{\rm vac}$ decouples sectors with different $M'$ quantum number, and has a tridiagonal structure within each sector, changing $J$ by at most one unit.

The contribution to the diagonal elements of the Hamiltonian follows from a corollary of the Wigner--Eckart theorem, the \emph{projection theorem}, which asserts that
\begin{equation}
    \bra{JM'} \hat{D}_k \ket{JM}
    = \frac{\bra{JM'} \hat{\vec{S}}_{\rm tot} \cdot \hat{\vec{D}} \ket{JM}}{J(J+1)}
       \bra{JM'} \hat{S}_{{\rm tot},k}' \ket{JM} .
    \label{eq:projection-theorem}
\end{equation}
Using $\vec S_{\rm{tot}} \cdot \vec D = \vec S_1^2 - \vec S_2^2$, \cref{eq:projection-theorem} leads to
\begin{align}
    \delta\Delta \bra{JM} \hat{D}_z \ket{JM}
      = \delta\Delta \frac{s_1(s_1+1) - s_2(s_2+1)}{J(J+1)} M.
    \label{eq:me-bipolar-vac-D}
\end{align}
Note that in the specific case $s_1 = s_2$, this contribution vanishes.

%-----------------------------------------------------------------------------
\subsection{Vacuum Term -- Off-Diagonal Part}
%-----------------------------------------------------------------------------

Concerning the off-diagonal term $\delta\Delta \, \bra{J-1,M} \hat{D}_z \ket{JM} = \delta\Delta \, \bra{JM} \hat{D}_z \ket{J-1,M}$, we first note that, since \smash{$\bra{J-1,M} \hat{\vec{S}}_{\rm tot} \ket{JM} = 0$}, it follows that \smash{$\bra{J-1,M} \hat{\vec{S}}_{2} \ket{JM} = -\bra{J-1,M} \hat{\vec{S}}_{1}\ket{JM}$}. Therefore,
\begin{align}
    \delta\Delta \, \bra{J-1,M} \hat{D}_z \ket{JM}
    = 2 \, \delta\Delta \, \bra{J-1,M} \hat{S}_{1,z} \ket{JM}.
    \label{eq:d-to-s1}
\end{align}
To evaluate this further, we decompose $\ket{J-1,M}$ and $\ket{JM}$ into their component spins using the Clebsch--Gordan coefficients:
\begin{widetext}
\begin{align}
    \bra{J-1,M} \hat{S}_{1,z} \ket{JM}
    &= \sum_{m_1,m_2,m_1',m_2'}
          \sprod{s_1 m_1 s_2 m_2}{J-1,M}
          \sprod{s_1 m_1' s_2 m_2'}{JM}
          \bra{s_1 m_1 s_2 m_2} \hat{S}_{1,z} \ket{s_1 m_1' s_2 m_2'}           \notag \\
    &= \sum_{m_1,m_2} m_1 \,
          \sprod{s_1 m_1 s_2 m_2}{J-1,M}
          \sprod{s_1 m_1 s_2 m_2}{JM}  \notag \\
    &= \sqrt{2J-1} \sqrt{2J+1}
          \sum_{m_1,m_2} m_1 \,
          \begin{pmatrix}
              s_1 & s_2 & J-1 \\
              m_1 & m_2 & -M
          \end{pmatrix}
          \begin{pmatrix}
              s_1 & s_2 & J \\
              m_1 & m_2 & -M
          \end{pmatrix} .
\end{align}
We now employ the recursion relation for the $3j$-symbols \cite{NIST:DLMF},
\begin{align}
    &(2j_1+1) \big[ j_2(j_2+1) - j_3(j_3+1) m_1 - j_1(j_1+1)(m_3-m_2) \big]
    \begin{pmatrix}
        j_1 & j_2 & j_3 \\
        m_1 & m_2 & m_3
    \end{pmatrix} \notag\\
    &\qquad
    = (j_1+1)
        \sqrt{j_1^2-(j_2-j_3)^2}
        \sqrt{(j_2+j_3+1)^2-j_1^2}
        \sqrt{j_1^2-m_1^2}
        \begin{pmatrix}
            j_1-1 & j_2 & j_3 \\
            m_1 & m_2 & m_3
        \end{pmatrix} \notag\\
    &\qquad
    + j_1 \sqrt{(j_1+1)^2-(j_2-j_3)^2}
        \sqrt{(j_2+j_3+1)^2-(j_1+1)^2}
        \sqrt{(j_1+1)^2-m_1^2}
        \begin{pmatrix}
            j_1+1 & j_2 & j_3 \\
            m_1 & m_2 & m_3
        \end{pmatrix}.
\end{align}
Using the symmetry properties of the $3j$-symbols to permute the columns, setting $j_1=s_1$, $j_2=s_2$, $j_3=J$, and $m_3=-M$, and isolating the term containing an $m_1$ prefactor, we obtain
\begin{align}
    &(2J+1) \, J(J+1) \, m_1
    \begin{pmatrix}
        s_1 & s_2 & J \\
        m_1 & m_2 & -M
    \end{pmatrix}
    = (J+1) \sqrt{J^2-(s_1-s_2)^2}
              \sqrt{(s_1+s_2+1)^2-J^2}
              \sqrt{J^2-M^2}
        \begin{pmatrix}
            s_1 & s_2 & J-1 \\
            m_1 & m_2 & -M
        \end{pmatrix} \notag\\
    &\qquad
    - (2J+1) \big[ s_1(s_1+1) + s_2(s_2+1) M
        - J(J+1) m_2 \big]
        \begin{pmatrix}
            s_1 & s_2 & J \\
            m_1 & m_2 & -M
        \end{pmatrix} \notag\\
    &\qquad
    + J \sqrt{(J+1)^2-(s_1-s_2)^2}
          \sqrt{(s_1+s_2+1)^2-(J+1)^2}
          \sqrt{(J+1)^2-M^2}
        \begin{pmatrix}
            s_1 & s_2 & J+1 \\
            m_1 & m_2 & -M
        \end{pmatrix}.
\end{align}
We multiply this expression by \smash{$\sqrt{2J-1} \sqrt{2J+1} \begin{pmatrix} s_1 & s_2 & J-1 \\ m_1 & m_2 & -M \end{pmatrix}$}, sum over $m_1$, $m_2$, and use the orthogonality properties of the $3j$-symbols:
\begin{multline}
    \sum_{m_1,m_2} \sqrt{2J-1} \sqrt{2J+1} \, m_1
        \begin{pmatrix}
            s_1 & s_2 & J-1 \\
            m_1 & m_2 & -M
        \end{pmatrix}
        \begin{pmatrix}
            s_1 & s_2 & J \\
            m_1 & m_2 & -M
        \end{pmatrix} \\
    = \frac{\sqrt{2J-1} \sqrt{2J+1}}
           {(2J+1) J(J+1)(2J-1)} (J+1)
        \sqrt{J^2-(s_1-s_2)^2}
        \sqrt{(s_1+s_2+1)^2-J^2}
        \sqrt{J^2-M^2} \\
    + \sum_{m_1,m_2} \sqrt{2J-1}\sqrt{2J+1} \, m_2
        \begin{pmatrix}
            s_1 & s_2 & J-1 \\
            m_1 & m_2 & -M
        \end{pmatrix}
        \begin{pmatrix}
            s_1 & s_2 & J \\
            m_1 & m_2 & -M
        \end{pmatrix} .
\end{multline}
In the last term on the right-hand side, we can relabel $s_1 \leftrightarrow s_2$ and $m_1 \leftrightarrow m_2$ and use the symmetry properties of the $3j$-symbols again to see that this term is identical, up to a sign, to the left-hand side and can therefore be combined with the latter. Our final result is then
\begin{align}
    \delta\Delta \, \bra{J-1,M} \hat{D}_z \ket{JM}
    = \delta\Delta \frac{\sqrt{J^2-M^2}}{J \sqrt{4J^2 - 1}}
        \sqrt{J^2-(s_1-s_2)^2}
        \sqrt{(s_1+s_2+1)^2-J^2}.
    \label{eq:me-bipolar-vac-offdiag}
\end{align}

%-----------------------------------------------------------------------------
\subsection{Summary}
%-----------------------------------------------------------------------------

Putting together the results derived in this section, in particular \cref{eq:me-bipolar-int,eq:me-bipolar-vac-D,eq:me-bipolar-vac-S,eq:me-bipolar-vac-offdiag}, the Hamiltonian matrix elements take the form
\begin{align}
    \bra{JM} \hat{H} \ket{JM}
    &= 2J_{12}\left[J(J+1)-s_1(s_1+1) - s_2(s_2+1)\right]
     + \bar\Delta M
     + \delta\Delta \frac{s_1(s_1+1) - s_2(s_2+1)}{J(J+1)} M,
     \\[6pt]
    \bra{J-1,M} \hat{H} \ket{JM}
    &= \delta\Delta \frac{\sqrt{J^2-M^2}}{J \sqrt{4J^2 - 1}}
        \sqrt{J^2-(s_1-s_2)^2}
        \sqrt{(s_1+s_2+1)^2-J^2}.
\end{align}
\end{widetext}
For the bipolar system with equal number of neutrinos and antineutrinos ($s \equiv s_1 = s_2 = \frac{n}{2}$ and $\Delta_1 = -\Delta _2$, and therefore $\bar\Delta = 0$), these expressions reduce to \cref{eq:H-bipolar-diag-final,eq:H-bipolar-offdiag-final}.

%=============================================================================
\section{Adaptive Classical Algorithm}
\label{sec:adaptive-classical}
%=============================================================================

The algorithm in~\cref{sec:bipolar-algo} achieves high performance by truncating the initial state to the most relevant $M'$ sectors (keeping in mind that $M'$ is a constant of motion), which introduces a constant error. Additional truncations are possible during time evolution at the cost of introducing additional errors in each step of the computation. This section describes the algorithm in more detail.

Within each $M'$ block, the algorithm operates on a set of states $\ket{JM'}$ that form a contiguous chain in $J$, $J \in [J_{\rm lo}, J_{\rm hi}]$, which is a subset of the full $J$ chain. At $t=0$, $J_{\rm lo}$ and $J_{\rm hi}$ in each block are chosen in the following way: the algorithm ranks the overlap of the initial state with each $\ket{JM}$ state. The state with the largest overlap is selected first, followed by the one with the next-to-largest overlap, and so on. This is repeated until the total amplitude of the left-out states is at most at a configured tolerance $\epsilon_J$. Then, within each $M'$ block, the retained $J$ states are closed into one contiguous interval by choosing $J_{\rm lo}$ and $J_{\rm hi}$ as the lowest and highest retained $J$ values, respectively, and keeping all states in between.

A runtime error estimation is used to dynamically expand the active window $[J_{\rm lo}, J_{\rm hi}]$ within each of the retained $M'$ blocks. The windows are only ever enlarged and never shrunk.  Since $\hat H$ is tridiagonal in $J$ within each $M'$ block, amplitude can leave an active $J$ window only through its two edges, via the off-diagonal matrix elements. The error is therefore monitored via the amplitude flux through the two edges. Let $\ket{\tilde\Psi(t)}$ denote the truncated state that the algorithm actually propagates, with its basis being the states within the active window. Denote the amplitudes of a particular basis state $\ket{JM'}$ as $\tilde\Psi_{JM'}(t) \equiv \sprod{JM'}{\smash{\tilde\Psi(t)}}$. During a time step $\delta t$, the amplitude transferred out of the upper edge of the active window to the first excluded state $\ket{J_{\rm hi}+1, M'}$ is approximately
\begin{align}
  \delta t \, \big| \bra{J_{\rm hi}+1, M'} \hat H \ket{J_{\rm hi}, M'} \big| \, \big| \tilde\Psi_{J_{\rm hi} M'}(t) \big| ,
\end{align}
where the matrix element is given by~\cref{eq:H-bipolar-offdiag-final}. The lower edge is treated in the same way. Before each time step, any window whose amplitude flux across the edge exceeds a preset threshold ($\varepsilon_{\rm thr} = 10^{-9}$ for the results in this paper) is extended on that side. The added states are initialized with zero amplitudes. Note that the window expansion threshold is not an error tolerance setting, as global error budget applies and will be explained below.

For a full error estimation, let $\hat P$ be the projector onto the active window of all retained $M'$ blocks, and $\hat Q \equiv 1 - \hat P$. We compare the exact state $\ket{\Psi(t)}$ with the truncated state $\ket{\tilde\Psi(t)}$. The latter starts from the normalized projection of the initial state onto the retained $M'$ sectors and $J$ windows and is evolved under $\hat P \hat H \hat P$. 
For the unitary time evolution, over any interval $[t_0, t]$ over which $\hat P$ is fixed, 
\begin{align}
  \big\| \ket{\Psi(t)} - \ket{\tilde\Psi(t)} \big\|
  \leq \big\| &\ket{\Psi(t_0)} - \ket{\tilde\Psi(t_0)} \big\| \nonumber \\
  &+ \int_{t_0}^{t} \! \mathrm{d}t' \, \big\| \hat Q \hat H \hat P \ket{\tilde\Psi(t')} \big\| .
\end{align}
The error due to leakage of amplitude outside the active $J$ window, $\hat Q \hat H \hat P \ket{\tilde\Psi}$, is supported on the two omitted states adjacent to each window. Therefore,
\begin{align}
    \eta^2(t) &\equiv \big\| \hat Q \hat H \hat P \ket{\tilde\Psi(t)} \big\|^2 
                            \nonumber \\[6pt]
    &= \sum_{M'} \Big( \big| \bra{J_{\rm lo}-1, M'} \hat H \ket{J_{\rm lo}, M'} 
                 \, \tilde\Psi_{J_{\rm lo} M'} \big|^2 \nonumber \\
    &\qquad + \big| \bra{J_{\rm hi}+1, M'} \hat H \ket{J_{\rm hi}, M'}
                 \, \tilde\Psi_{J_{\rm hi} M'} \big|^2 \Big).
\end{align}
The solver can then accumulates the leakage error \smash{$\varepsilon_{\rm leak}(t) = \int_0^t \mathrm{d}t' \, \eta(t')$} by an endpoint-trapezoid estimation.

In this work, for $N = 3$ to $N= 40\,000$, the error remained below \num{2.1e-5} for both toy scenarios. In the two largest runs we carried out, we obtained an accumulated error of \num{1.8e-3} for the $\nu_e$--$\nu_\mu$ system at $N = 1\,300\,000$ with an expansion threshold of $\varepsilon_{\rm thr} = \num{e-9}$, and \num{9.5e-7} for the neutrino--antineutrino system at $N = 1\,240\,000$ with $\varepsilon_{\rm thr} = \num{e-17}$.

%=============================================================================
\bibliography{refs.bib}
%=============================================================================

\end{document}